\documentclass[a4paper,12pt,reqno]{amsart}
\pdfoutput=1

\usepackage{a4}

\usepackage[british]{babel}
\usepackage{float}
\usepackage{multirow}

\usepackage[T1]{fontenc}
\usepackage[utf8]{inputenc}

\usepackage{tikz_graphs}

\usepackage{subcaption}

\newtheorem{proposition}{Proposition}
\newtheorem{example}{Example}
\newtheorem{remark}{Remark}
\newtheorem{convention}{Convention}

\DeclareMathOperator{\Res}{Res}
\DeclareMathOperator{\Tr}{Tr}

\allowdisplaybreaks[4]

\begin{document}

\title[Two Approaches For a Perturbative Expansion in BTR]{Two Approaches For a Perturbative Expansion in Blobbed Topological Recursion}

\author[Jakob Lindner]{Jakob Lindner}
 
\address{\textsuperscript{1}Fachbereich Mathematik, Universit\"at Hamburg, 20146 Hamburg, Germany \hfill \newline
{\itshape e-mail:} \normalfont
\texttt{jakob.lindner@studium.uni-hamburg.de}}



\begin{abstract}
In this paper we continue the perturbative analysis of the quartic Kontsevich model. We investigate meromorphic functions $\Omega^{(0)}_m$ with $m=1,2$, that obey blobbed topological recursion. We calculate their expansions and check their equivalence to sums of ribbon graph weights, which are obtained with common methods of perturbation theory in QFT, up to fifth order in the coupling using Mathematica.

Furthermore, we provide a catalog of permutation pairs $(\alpha,\sigma)$, which encode all 5660 vacuum ribbon graphs that contribute to the free energy $\mathcal{F}^{(g)}$ with genus $g\geq 0$ up to fifth order and begin to expand upon the used methods to also consider ribbon graphs of general correlation functions $G_{\dots}$. This is a first step towards automation of the calculation of ribbon graph expansions in the quartic Kontsevich model.   
\end{abstract}

\maketitle
\markboth{\hfill\textsc\shortauthors}{\textsc{{Two Approaches For a Perturbative Expansion in BTR}\hfill}}

\maketitle
\flushbottom

\section{Introduction}
This paper continues the perturbative analysis of the quartic Kontsevich model (QKM), which was initiated in \cite{Branahl:2021uxs}. The QKM is a matrix model analogous to the famous Kontsevich model \cite{Kontsevich:1992ti}, but with a quartic instead of a cubic interaction term, and recently was identified to be a prime example of blobbed topological recursion (BTR) \cite{Branahl:2020yru,Hock:2021yru,Branahl:2021}, an extension \cite{Borot:2015hna} of the universal structure of topological recursion (TR) of Eynard and Orantin \cite{Eynard:2007kz}.

The QKM is closely related to a scalar quantum field theory with quartic self interaction on a noncommutative geometry, so a perturbative analysis of its meromorphic functions $\Omega_m^{(g)}(z_1,...,z_m)$, which obey blobbed topological recursion, is of great interest. These functions are expanded as a power series in a parameter $\lambda$ (the coupling constant in QFT jargon) and $N$ eigenvalues $(E_1,\dots,E_N)$ of the external field/matrix resulting in a sum of ribbon graphs (fat Feynman graphs, maps), a ribbon graph expansion.

We continue the analysis in the present paper with support of Mathematica programs by implementing two complementary approaches, which were introduced in \cite{Branahl:2021uxs}, to calculate these ribbon graph expansions. The equivalence of their result will be proven in each order up to $\lambda^5$.
Ribbon graph expansions of $\Omega_m^{(g)}(z_1,...,z_m)$, denoted by $\Omega^{(g)}_{q_1,\dots,q_m}$, are obtained at particular points in the complex variables $z_i=\varepsilon_i$, so one can interpret the meromorphic functions as complexified objects, analytically continued around $z_i=\varepsilon_i$ with condition $R(\varepsilon_i)=E_i$, where $R(z)$ is a variable transform that also determines the initial data of the topological recursion. 

The first approach involves a boundary creation operator (BCO) $\hat{T}$ that is repeatedly applied to the free energy $\mathcal{F}^{(g)}$ to directly obtain the ribbon graph expansions $\Omega^{(g)}_{q_1,\dots,q_m}$. Its difficulty lies in knowing the graph expansion of the free energy, so we implement methods to automatically calculate the contributing ribbon graphs, resulting in a catalog of 5660 graphs up to fifth order.
The second approach is based on the meromorphic functions $\Omega_m^{(g)}(z_1,...,z_m)$, evaluated at the special points $z_i=\varepsilon_i$. Expanding the occurring $\varepsilon_i$ and $R'(\varepsilon_i)$ as power series in $\lambda$, we obtain the ribbon graph expansion in a different partial fraction decomposition.
We check the equivalence of the results of both methods. In this paper we will focus on the planar case $g=0$ with $m=1,2$ up to order $\lambda^5$.

\smallskip 
The following section \ref{ch:Model} will introduce the QKM, briefly discusses the perturbative expansion of its cumulants, gives a short overview of TR and BTR and properly introduces the two approaches for obtaining a perturbative expansion of the $\Omega^{(g)}_m$. Section \ref{ch:vacuumGraphs} discusses the free energy $\mathcal{F}^{(g)}$, which is of importance for the first approach. We determine its vacuum ribbon graphs by using an encoding with permutations. The result is a catalog of encodings and properties of all vacuum ribbon graphs up to fifth order. Additionally, we briefly generalize this method to ribbon graphs with boundaries. Some examples from said catalog are given in appendix \ref{ch:B_examplesCatalog}. Section \ref{ch:PerturbativeExpansion} begins with a discussion of the BCO and its action on ribbon graphs. This is done both in the language of ribbon graph weights using pictures and encodings with permutations. More examples are given in appendix \ref{ch:appendixPictures}. We then list in section \ref{ch:PerturbativeExpansion} the numbers of individual graphs and weights that contribute to the perturbative expansions of the $\Omega^{(g)}_m$. Finally, we state the results of the comparison of the two approaches. Their expansions are equivalent up to fifth order in the cases of $\Omega^{(0)}_1$ and $\Omega^{(0)}_2$. We conclude with section \ref{ch:conclusion} and give an outlook on what remains. All Mathematica notebooks and the catalog are provided at \cite{Zenodo}. Appendix \ref{ch:A_remarksImplementation} gives a brief overview of their functionality. 

We refer to \cite{Branahl:2021} for a more detailed introduction to the model, the progress achieved in recent years and the relation of the QKM to other models like the Kontsevich model \cite{Kontsevich:1992ti} and the Hermitian 1- and 2-matrix model \cite{Eynard:2016yaa,Chekhov:2006vd}. As already mentioned, this paper extends some of the results obtained in \cite{Branahl:2021uxs} and hence will refer to and cite from the latter on multiple occasions. 

\centerline{\sc Acknowledgements}
\noindent I would like to thank Johannes Branahl and Raimar Wulkenhaar for numerous helpful comments.
\smallskip  

{\center \small \tableofcontents}

\section{The setup}\label{ch:Model}

\subsection{Quartic Kontsevich model}\label{ch:QKM}
Let $H_N$ denote the real vector space of self-adjoint $N \times N$-matrices, $H'_N$ will be its dual space and $(E_1,\dots,E_N)$ are pairwise different positive real numbers denoting the energy eigenvalues of a matrix $M = M^* = \sum_{k,l = 1}^N M_{kl}e_{kl} \in H_N$. By $(e_{kl})$ we denote the standard matrix basis in the complexification of $H_N$. The quantum scalar fields $\Phi$ are noncommutative random variables distributed on $H'_N$ according to a Gaussian measure $d\mu_{E,0}$ with covariance
\begin{equation}\label{eq:def_covariance}
	\int_{H'_N} d\mu_{E,0}(\Phi) \Phi(e_{jk}) \Phi(e_{lm}) = \frac{\delta_{jm} \delta_{kl}}{N(E_j+E_l)} \, .
\end{equation}
The Gaussian measure is deformed by a quartic potential to a measure 
\begin{align*}
	d\mu_{E,\lambda}(\Phi) &= \frac{1}{\mathcal{Z}} \exp \Big(-\frac{\lambda N}{4} \Tr(\Phi^4)\Big) d\mu_{E,0}(\Phi) \, , \label{eq:def_measure} \\
		&\quad \text{where} \quad \mathcal{Z} = \int_{H'_N} \exp\Big(-\frac{\lambda N}{4} \Tr(\Phi^4)\Big) d\mu_{E,0}(\Phi) \, . \nonumber
\end{align*}
Here $\lambda$ is the coupling of the interaction. As mentioned before, this deformation is different from the Kontsevich model \cite{Kontsevich:1992ti}, where the measure is deformed by a cubic potential $\Tr(\Phi^3)$ instead of a quartic $\Tr(\Phi^4)$. Partially differentiating the Fourier transform of the measure
\begin{equation}\label{eq:def_generatingFunctional}
	\mathcal{Z}(M) = \int_{H'_N} d\mu_{E,\lambda}(\Phi) e^{i\Phi(M)} \, ,
\end{equation}
also called the \emph{partition function}, with respect to matrix entries $M_{kl}$ gives expectation values of variable products called \emph{moments}
\begin{equation*}
	\left< e_{k_1l_1}\dots e_{k_nl_n} \right> = \int_{H'_N} d\mu_{E,\lambda}(\Phi) \Phi(e_{k_1l_1}) \cdots \Phi(e_{k_nl_n}) = \frac{1}{i^n} \frac{\partial^n \mathcal{Z}(M)}{\partial M_{k_1l_1} \cdots \partial M_{k_nl_n}} \bigg|_{M=0} \, .
\end{equation*}
\emph{Cumulants} can be calculated in a similar way using the partition function \eqref{eq:def_generatingFunctional} with
\begin{equation*}\label{eq:def_cumulants}
	\left< e_{k_1l_1}\dots e_{k_nl_n} \right>_c = \frac{1}{i^n} \frac{\partial^n \log \mathcal{Z}(M)}{\partial M_{k_1l_1} \cdots \partial M_{k_nl_n}} \bigg|_{M=0} \, .
\end{equation*} 
The moments can be decomposed into cumulants according to \cite{Schurmann:2021mzu}
\begin{equation*}
	\Big< \prod_{i=1}^n e_{k_il_i} \Big> = \sum_{\substack{\text{partitions} \\ \pi \, \text{of} \, \{1,\dots,n\}}} \prod_{\text{blocks} \, \beta \in \pi} \Big< \prod_{i \in \beta} e_{k_il_i} \Big>_c \,.
\end{equation*}
In the case of a quartic potential only moments and cumulants with even $n$ are nonvanishing, so every block $\beta$ in the decomposition has to be of even length. The covariance \eqref{eq:def_covariance} of the Gaussian measure and the invariance of the trace under cyclic permutations give additional constraints on the cumulants. A cumulant is only nonzero if its indices $(l_1,\dots,l_n) = (k_{\sigma(1)},\dots,k_{\sigma(n)})$ are a permutation $\sigma \in S_n$ of $(k_1,\dots,k_n)$ with $b$ cycles, where $S_n$ is the symmetric group. Under the assumption of pairwise different $k_i^j$ it is connected to the $(n_1+\dots+n_b)$-point function $G_{\dots}$
\begin{equation}\label{eq:def_G}
	N^{n_1+\cdots+n_b} \big<(e_{k_1^1k_2^1}e_{k_2^1k_3^1} \cdots e_{k_{n_1}^1k_1^1}) \cdots  (e_{k_1^bk_2^b}e_{k_2^bk_3^b} \cdots e_{k_{n_b}^bk_1^b})\big>_c = N^{2-b} G_{|k_1^1 \dots k_{n_1}^1| \dots |k_1^b \dots k_{n_b}^b|} \, .
\end{equation}
Note the change in notation: $k_i^j$ corresponds to the $i$-th index of the $b$-th cycle of the permutation $\sigma$ and it should not be confused with the previous $k_i$. 
The $G_{\dots}$, also called \textit{correlation functions}, can be further expanded $G_{\dots} = \sum_{g=0}^{\infty} N^{-2g} G_{\dots}^{(g)}$ into correlation functions of genus $g$. 

\subsection{Perturbative series}\label{ch:PertSeries}

These $G_{\dots}^{(g)}$ are closely related to connected \textit{ribbon graphs} with four-valent vertices, where each graph can be embedded into a bordered Riemann surface $\mathcal{C}_{g,b}$ with genus $g\geq0$, number $b\geq1$ of boundary components and an \textit{Euler characteristic} $\chi = 2-2g-b$ (we refer to \cite{Branahl:2021uxs} for the details of this connection). 

Likewise, the moments, which will not be considered here, can possibly correspond to disconnected ribbon graphs, which are not embedded into a single, but multiple Riemann surfaces, one surface for each connected subgraph of the disconnected graph. A graph is \emph{connected} if every vertex is connected to every other vertex via an alternating chain of ribbons and vertices, otherwise it is called a \emph{disconnected} graph. 

 Ribbon graphs, also called \textit{fatgraphs}, give a visual interpretation of the correlation functions. Their underlying notation dates back to t'Hooft \cite{Hooft:1974}, who used them in the context of a theory for strong interactions. Unlike normal Feynman graphs with edges that are drawn as a single line, ribbon graphs have edges with two lines, called \emph{strands}. Together they form a \emph{ribbon}. Each strand has an index associated with it.
 
 The connection between correlation functions $G^{(g)}_{|p_1^1...p_{n_1}^1|...|p_1^b...p_{n_b}^b|}$ and the set of graphs $\mathfrak{G}^{g,v}_{|p_1^1...p_{n_1}^1|...|p_1^b...p_{n_b}^b|}$, which are embedded into an oriented surface $\mathcal{C}_{g,b}$ and have $v$ four-valent vertices, is made explicit in
\begin{equation}
	G^{(g)}_{|p_1^1...p_{n_1}^1|...|p_1^b...p_{n_b}^b|} =\sum_{v=0}^\infty \, \sum_{\Gamma \in \mathfrak{G}^{g,v}_{|p_1^1...p_{n_1}^1|...|p_1^b...p_{n_b}^b|}} \varpi(\Gamma) \, ,\label{eq:sum_graphs}
\end{equation}
where $\varpi(\Gamma)$ is a \emph{weight} associated with a single graph $\Gamma$.
\begin{proposition}[Feynman Rules \cite{Branahl:2021uxs}]\label{prop:FeynmanRules}
The weight $\varpi(\Gamma)$ of each graph $\Gamma$ can be determined with the following \emph{Feynman rules}
	\begin{itemize}
		\item every loop (closed strand) with label $k$ contributes a factor $\frac{1}{N}$ and a summation $\sum_{k=1}^N$
		\item every $4$-valent vertex contributes a factor $-\lambda$
		\item every ribbon with strands labelled $p,q$ contributes a factor $\frac{1}{E_p+E_q}$.
	\end{itemize}
	\noindent
	The product of all factors with the summation over all loops gives the weight.
\end{proposition}
The Euler characteristic $\chi$ of the surface, in which the graph is embedded, can also be determined by the number of four-valent vertices $v$, ribbons $r$, one-valent vertices $n$ and loops $l$ \begin{equation}\label{eq:Euler}
	\chi = 2 - 2g - b \stackrel{!}{=} v - r + n + l \, .
\end{equation}

\begin{example}
The $2$-point function $G^{(1)}_{|ab|}$ with genus $g=1$ has six graphs at order $\lambda^2$, which are displayed in figure \ref{fig:2point_g1_lam2}. There are no contributions at zeroth or first order. According to the Feynman rules above the sum of the weights is given by
\begin{align}
	G_{|ab|}^{(1)} &= \frac{(-\lambda)^2}{(E_a+E_b)^2} \bigg(\frac{1}{4E_aE_b(E_a+E_b)} + \frac{1}{(2E_a)^3} + \frac{1}{(2E_b)^3} \nonumber \\
	&\qquad + \frac{1}{(E_a+E_b)^3} + \frac{1}{(2E_a)^2(E_a+E_b)} + \frac{1}{(2E_b)^2(E_a+E_b)} \bigg) + \mathcal{O}[\lambda^3] \, . \label{eq:example1}
\end{align} 
\begin{figure}[H]
	\centering
		\def\svgwidth{0.7\columnwidth}
		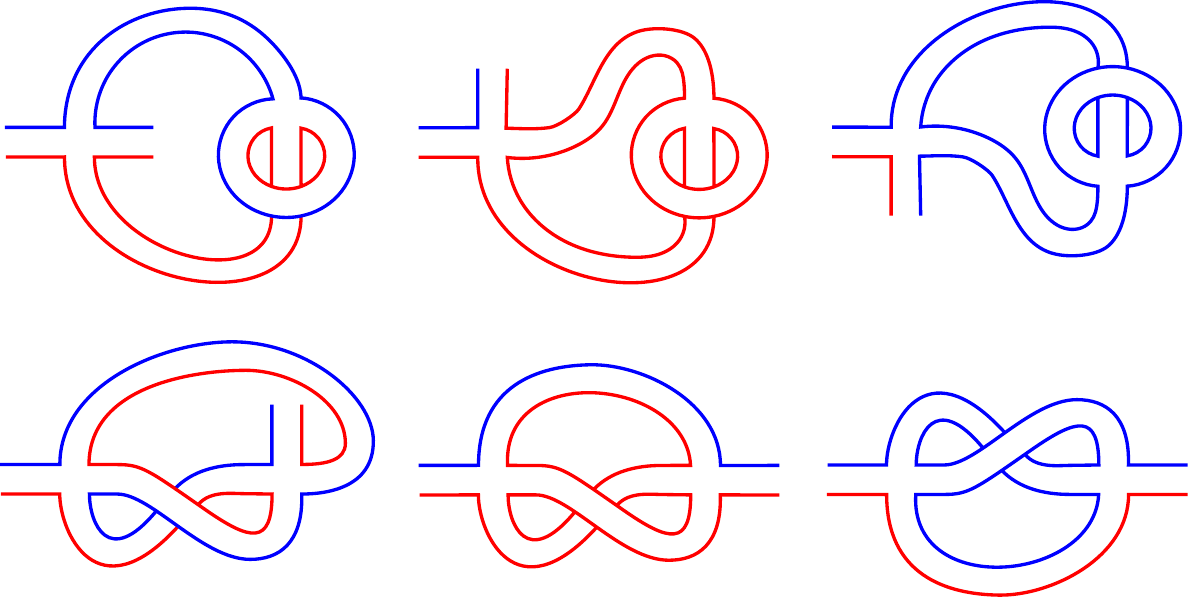
	\caption{All six graphs that contribute to $G^{(1)}_{|ab|}$ in second order. Red strands have label $a$ and blue strands $b$.}
	\label{fig:2point_g1_lam2}
\end{figure}

\end{example}
Similar to \eqref{eq:sum_graphs}, which defines correlation functions $G_{\dots}^{(g)}$ as a sum over the weights $\varpi(\Gamma)$ of ribbon graphs $\Gamma \in \mathfrak{G}^{g,v}_{|p_1^1...p_{n_1}^1|...|p_1^b...p_{n_b}^b|}$, we introduce the \emph{free energy} of arbitrary genus $g$
\begin{equation}
	\mathcal{F}^{(g)}= -\frac{\delta_{g,0}}{2N^2} \sum_{j,k=1}^N \log(E_j+E_k) +\sum_{v=1}^\infty  \sum_{\Gamma_0\in\mathfrak{G}^{g,v}_\emptyset}\frac{\varpi(\Gamma_0)}{|\mathrm{Aut}(\Gamma_0)|} \,. \label{eq:FreeEnergyGeneral}
\end{equation}
In this case only graphs $\Gamma_0$, which can be drawn on Riemann surfaces without boundary components $(b=0)$ and thus belong to the set $\mathfrak{G}^{g,v}_\emptyset$, are considered. They are called \emph{vacuum graphs} \cite{Brezin:1977sv}. Here $|\text{Aut}(\Gamma_0)|$ denotes the order of the automorphism group of the graph. Roughly speaking, an automorphism of a graph is a permutation of edges and vertices under which the graph is kept invariant. Every graph, whose corresponding Riemann surface has at least one boundary $b\geq1$, has a trivial automorphism group $\text{Aut}=\{\text{id}\}$ and thus is not mentioned in \eqref{eq:sum_graphs}.

\begin{example}\label{exmp:freeEnergy}
The graphs of $\mathcal{F}^{(0)}$ up to order $\lambda^2$ were introduced in \cite{Branahl:2021uxs} and are recited in figure \ref{fig:Lambda012} in simplified notation. Two strands of ribbons of vacuum graphs on the sphere $(g=0)$ always have distinct indices and thus can be drawn as a single line. Such a graph partitions the sphere into connected surfaces, which are homeomorphic to disks. Each connected surface is then associated with an index of a closed strand. 

The order of the automorphism groups of each graph is also given in figure \ref{fig:Lambda012}. The graph on the right is invariant under a swap of its two vertices and a simultaneous rotation of both vertices, thus resulting in $|\text{Aut}|=2\cdot4= 8$. The single graph in order $\lambda^1$ is only invariant under a rotation of its vertex by $\pi$ and a subsequent swap of the ribbons, giving $|\text{Aut}|= 2$. At order $\lambda^0$ there is only one graph, whose weight cannot be determined by the Feynman rules of remark \ref{prop:FeynmanRules}. It corresponds to the first term in \eqref{eq:FreeEnergyGeneral}. The free energy $\mathcal{F}^{(0)}$ with genus $g=0$ can thus be written explicitly up to second order
\begin{align}
	\mathcal{F}^{(0)}
		&= \frac{-1}{2N^2} \sum_{j,k=1}^N \log(E_j{+}E_k) + \frac{(-\lambda)}{N^3} \sum_{j,k,l=1}^N \frac{1}{2(E_j{+}E_k)(E_j{+}E_l)} \nonumber \\
		&\quad + \frac{(-\lambda)^2}{N^4} \sum_{j,k,l,m=1}^N \bigg(\frac{1}{8(E_j{+}E_k)(E_k{+}E_l)(E_l{+}E_m)(E_m{+}E_j)}  \nonumber \\
		&\quad + \frac{1}{2(E_j{+}E_k)(E_j{+}E_l)^2(E_j{+}E_m)} + \frac{1}{2(E_j{+}E_k)(E_j{+}E_l)^2(E_l{+}E_m)} \bigg) + \mathcal{O}[\lambda^3] \label{eq:free_energy} \,.
\end{align}

\begin{figure}[H]
	\centering
		\def\svgwidth{0.6\columnwidth}
		%% Creator: Inkscape 1.1.1 (c3084ef, 2021-09-22), www.inkscape.org
%% Accompanies image file 'Lambda012_new.pdf' (pdf, eps, ps)
\begingroup%
  \makeatletter%
  \providecommand\color[2][]{%
    \errmessage{(Inkscape) Color is used for the text in Inkscape, but the package 'color.sty' is not loaded}%
    \renewcommand\color[2][]{}%
  }%
  \providecommand\transparent[1]{%
    \errmessage{(Inkscape) Transparency is used (non-zero) for the text in Inkscape, but the package 'transparent.sty' is not loaded}%
    \renewcommand\transparent[1]{}%
  }%
  \providecommand\rotatebox[2]{#2}%
  \newcommand*\fsize{\dimexpr\f@size pt\relax}%
  \newcommand*\lineheight[1]{\fontsize{\fsize}{#1\fsize}\selectfont}%
  \ifx\svgwidth\undefined%
    \setlength{\unitlength}{368.24997541bp}%
    \ifx\svgscale\undefined%
      \relax%
    \else%
      \setlength{\unitlength}{\unitlength * \real{\svgscale}}%
    \fi%
  \else%
    \setlength{\unitlength}{\svgwidth}%
  \fi%
  \global\let\svgwidth\undefined%
  \global\let\svgscale\undefined%
  \makeatother%
  \begin{picture}(1,0.42898207)%
    \lineheight{1}%
    \setlength\tabcolsep{0pt}%
    \put(0,0){\includegraphics[width=\unitlength,page=1]{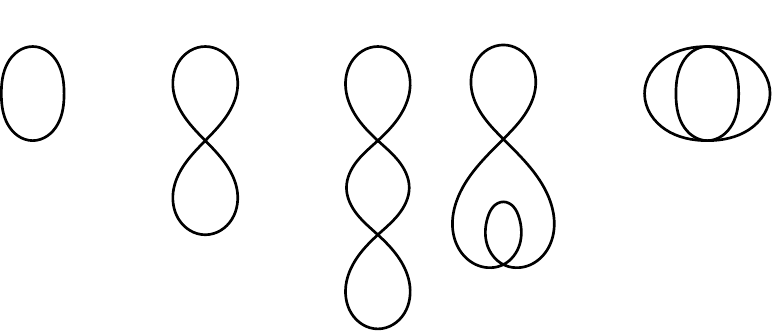}}%
    \put(0.255,0.40835033){\makebox(0,0)[lt]{\lineheight{1.25}\smash{\begin{tabular}[t]{l}$2$\end{tabular}}}}%
    \put(0.56,0.40835033){\makebox(0,0)[lt]{\lineheight{1.25}\smash{\begin{tabular}[t]{l}$2$\end{tabular}}}}%
    \put(0.905,0.40835033){\makebox(0,0)[lt]{\lineheight{1.25}\smash{\begin{tabular}[t]{l}$8$\end{tabular}}}}%
  \end{picture}%
\endgroup%

	\caption{The vacuum ribbon graphs of $\mathcal{F}^{(0)}$ in zeroth, first and second order ($\Gamma_0\in\mathfrak{G}^{0,v}_\emptyset$ with $v\in\{0,1,2\}$) with their order $|\text{Aut}(\Gamma_0)|$ of the automorphism group.}
	\label{fig:Lambda012}
\end{figure}
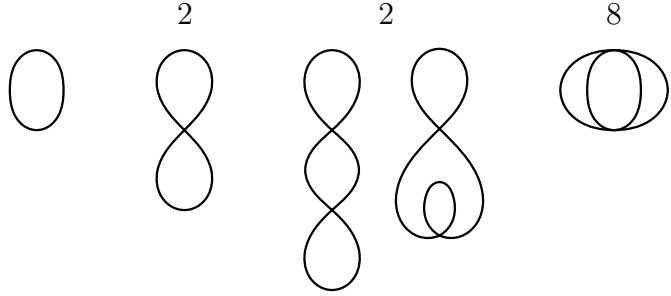  
\end{example}

Note that there are two types of indices occurring in \eqref{eq:example1} and \eqref{eq:free_energy}. The indices $a$ and $b$ of the energy eigenvalues appearing in the expansion of $G^{(1)}_{|ab|}$ are fixed, so they are called \emph{fixed indices}. In contrast, the indices of the free energy $\mathcal{F}^{(0)}$ are not fixed, as there is a summation over the whole set $(E_1,\dots,E_N)$ of eigenvalues. They are called \emph{free indices}.   

\begin{remark}
	We refer to \cite{Branahl:2021uxs} for additional examples. There, the graphs contributing in the first orders of the $2$-point function $G^{(0)}_{|ab|}$, the $(2+2)$-point function $G^{(0)}_{|ab|cd|}$ and the $4$-point function $G^{(0)}_{|abcd|}$ are displayed. All vacuum ribbon graphs of genus $g=1$ and their weights up to second order are given in \cite{branahl2021genus}.
\end{remark}
By repeatedly applying a \emph{boundary creation operator} (BCO) $\hat{T}_q=- N \frac{\partial}{\partial E_q}$ to the free energy we obtain a graph expansion
\begin{align}
	\Omega^{(g)}_{q_1,\dots,q_m} &= \hat T_{q_2} \dots \hat T_{q_m} \Omega^{(g)}_{q_1} + \frac{\delta_{g,0}\delta_{m,2}}{(E_{q_1}-E_{q_2})^2} \, ,\label{eq:generalDef_OmegaF} \\
	&\quad \text{with} \quad \Omega^{(g)}_{q_1} = \frac{1}{N} \sum_{k=1}^N G^{(g)}_{|q_1k|} + \frac{1}{N^2} G^{(g-1)}_{|q_1|q_1|} = \hat{T}_{q_1} \mathcal{F}^{(g)} \nonumber \, ,
\end{align}
of the meromorphic functions $\Omega_m^{(g)}$. This equation is the first of the two approaches that will be implemented below. 

\subsection{Blobbed topological recursion}\label{ch:BTR}

Over the last years the algebraic structure of the Quartic Kontsevich Model (QKM) was identified to be \emph{blobbed topological recursion} (BTR). This has been conjectured in \cite{Branahl:2020yru}, supported by some first evidence, and has been proved in \cite{Hock:2021yru} in the case of $g=0$. BTR is an extension of \emph{topological recursion} (TR) and was developed in 2015 by G. Borot and S. Shadrin \cite{Borot:2015hna} based on the framework of TR by B. Eynard and N. Orantin from 2007 \cite{Eynard:2007kz}.

TR gives a recursive definition of $\omega_{g,m}$, which are symmetric meromorphic $m$-forms ($m$-fold tensor products of 1-forms) on $\Sigma^m$, a product of $m$ Riemann surfaces $\Sigma$ (one-dimensional complex manifolds). Given some initial data $(\Sigma,\Sigma_0,x,$ $\omega_{0,1},B)$, which is called a \emph{spectral curve}, all differential forms of higher topology can be calculated from differential forms with lower topology using the recursion equation
	\begin{align}
		& \omega_{g,|I|+1}(I,z) \label{eq:TR_recursion} \\
  		& =\sum_{\beta_i} \Res\displaylimits_{q\to \beta_i} K_i(z,q)\bigg( \omega_{g-1,|I|+2}(I, q,\sigma_i(q)) +\hspace*{-1cm} \sum_{\substack{g_1+g_2=g\\ I_1\uplus I_2=I\\
   		(g_1,I_1)\neq (0,\emptyset)\neq (g_2,I_2)}} \hspace*{-1.1cm} \omega_{g_1,|I_1|+1}(I_1,q) \omega_{g_2,|I_2|+1}(I_2,\sigma_i(q))\!\bigg) \nonumber \, ,
	\end{align}
	where $I=\{z_1,\dots,z_m\}$ denotes the spectator variables besides $z$. Here $x:\Sigma \to \Sigma_0$ is a ramified covering of Riemann surfaces and has \emph{ramification points} $\beta_i$ that are defined by $x'(\beta_i)=0$.
	In the case of TR the $2$-form $\omega_{0,2}$ is equivalent to the \emph{Bergman kernel} $B(z,u)=\frac{dz \, du}{(z-u)^2}$. The \emph{recursion kernel} $K_i(z,q)$ is given with the initial data in the neighborhood of the ramification points $\beta_i$ as
	\begin{equation*}
		K_i(z,q) = \frac{1}{2} \frac{1}{\omega_{0,1}(q)-\omega_{0,1}(\sigma_i(q))} \int_q^{\sigma_i(q)} B(z,q') 
	\end{equation*} 
	and the $\sigma_i \neq \mathrm{id}$ are local (around $\beta_i$) \emph{Galois involutions} $x(q)=x(\sigma_i(q))$. Examples for research fields, where TR appears, are the Kontsevich model \cite{Kontsevich:1992ti,Eynard:2007kz}, Gromov-Witten theory \cite{Bouchard:2007ys}, the 1-matrix and 2-matrix model \cite{Chekhov:2006vd} and Hurwitz theory \cite{Bouchard:2007hi}. \\ 
	BTR is an extension of TR in the sense that the $m$-forms $\omega_{g,m}$ decompose
	\begin{align}
		\omega_{g,m}(I,z)=\mathcal{P}_z\omega_{g,m}(I,z)+\mathcal{H}_z\omega_{g,m}(I,z) \label{eq:BTR_recursion}
	\end{align}
	into a polar part $\mathcal{P}_z \omega_{g,m}$ that has poles (in a selected variable $z$) at the ramification points $\beta_i$ and a holomorphic part $\mathcal{H}_z \omega_{g,m}$ with poles at positions that are different from the $\beta_i$'s. The former part obeys the recursion equation \eqref{eq:TR_recursion} of TR and can be calculated from full $\omega_{g',m'}$ of lower topology given by the lhs of \eqref{eq:BTR_recursion}. For the latter, which is exclusive to BTR and contributes to the differential forms in every iteration step of \eqref{eq:TR_recursion}, there is currently no universal structure known. Thus one has to solve \emph{abstract loop equations} \cite{Borot:2013lpa} for each individual pair $(g,m)$ to determine the $\omega_{g,m}$. There is one exeption. In \cite{Hock:2021yru} a recursion formula, similar to \eqref{eq:TR_recursion}, was found for the planar ($g=0$) case of the QKM, making an iterative calculation of the $\omega_{0,m}$ possible.
	BTR receives its name from holomorphic parts (for $\chi = 2-2g-m < 0$) $\phi_{g,m}(z_1,...,z_{m-1},z)=\mathcal{H}_{z_1}...\mathcal{H}_{z_{m-1}}\mathcal{H}_z\omega_{g,m}(z_1...,z_{m-1},z)$ that are called \emph{blobs}.
	
The loop equations of the QKM have been found and solved in \cite{Branahl:2020yru} for the cases $(g,m) \in \{(0,2),(0,3),(0,4),(1,1)\}$. The spectral curve $(\Sigma,\Sigma_0,x,\omega_{0,1},B)$ of genus zero was identified with $x=R:\hat{\mathbb{C}} \to \hat{\mathbb{C}}$ ($\hat{\mathbb{C}}= \mathbb{C} \cup \{\infty\}$ is the \emph{Riemann sphere}), $B$ is the Bergman kernel and $\omega_{0,1}(z)=-R(-z)R'(z)dz$. In contrast to TR, where the $2$-form $\omega_{0,2}$ is equivalent to the Bergman kernel $B$, an additional blob $\phi_{0,2}$ has to be included in our case of BTR
	\begin{equation*}
		\omega_{0,2}(u,z) = B(u,z) + \phi_{0,2} = \frac{du\,dz}{(u-z)^2} + \frac{du\,dz}{(u+z)^2} \,.
	\end{equation*}
	The function $R$ in $\omega_{0,1}$ is meromorphic and was established in \cite{Grosse:2019jnv}
	\begin{equation}\label{eq:R}
 		R(z) = z-\frac{\lambda}{N} \sum_{k=1}^d \frac{r_k}{R'(\varepsilon_k)(\varepsilon_k+z)} 
 	\end{equation}
 	with $R(\varepsilon_k)=e_k$ and $R(\infty)=\infty$.  
 	Here, the $(e_1,\dots,e_d)$ are pairwise different eigenvalues with respective multiplicities $(r_1,\dots,r_d)$ with $\sum_{k=1}^d r_k = N$. Furthermore, a connection 
 	\begin{align}\label{eq:connection_omegas} 
  		\omega_{g,m}(z_1,\dots,z_m) 
			&:= \lambda^{2-2g-m} \Omega^{(g)}_m(z_1,\dots,z_m) \prod_{i=1}^m R'(z_i) dz_i
	\end{align}
 	between the meromorphic $m$-forms $\omega_{g,m}$ with $\chi = 2-2g-m < 0$ and meromorphic functions $\Omega^{(g)}_m$ was established. 	

\subsection{Two approaches}\label{ch:TwoApproaches}

We are interested in perturbative expansions of the functions $\Omega^{(g)}_m$ into formal power series in $\lambda$, where the $j$-th summand is a sum of weights that are related to (connected and possibly disconnected) graphs with $v=j$ vertices. 
Two different ways of calculating a specific order of this series were introduced in \cite{Branahl:2021uxs}. The first is the recursive definition given in \eqref{eq:generalDef_OmegaF} where the boundary creation operator $\hat{T}_q=- N \frac{\partial}{\partial E_q}$ is repeatedly applied to the free energy $\mathcal{F}^{(g)}$.

For the second approach we note that the $\Omega^{(g)}_m(z_1,\dots,z_m)$ in \eqref{eq:connection_omegas} are the complexifications of $\Omega^{(g)}_{q_1,\dots,q_m}$ in \eqref{eq:generalDef_OmegaF}. The latter can be obtained from the former by evaluation at points $z_i=\varepsilon_i \in \mathbb{R}$
 	\begin{equation*}
 		\Omega^{(g)}_m (\varepsilon_1,\dots,\varepsilon_m) = \Omega^{(g)}_{q_1,\dots,q_m} \, ,
 	\end{equation*}
 	where $\varepsilon_i$ is defined by the function $R(z)$ of the spectral curve. The $\varepsilon_q$ can be perturbatively expanded by inserting $\varepsilon_q$ into \eqref{eq:R} yielding 
 	\begin{equation}\label{eq:eps_exact}
 		\varepsilon_q = e_q + \frac{\lambda}{N} \sum_{n=1}^d \frac{r_n}{R'(\varepsilon_n)(\varepsilon_q+\varepsilon_n)} \, .
 	\end{equation}
 	Full expressions for the functions
 	\begin{align}
 		\Omega_1^{(0)}(z) &= -\frac{R(-z)+R(z)}{\lambda} - \frac{1}{N} \sum_{k=1}^d \frac{r_k}{R(\varepsilon_n)-R(z)} \label{eq:Omega1} \\
 		\Omega_2^{(0)}(u,z) &= \frac{1}{R'(u)R'(z)} \left( \frac{1}{(u-z)^2} + \frac{1}{(u+z)^2} \right) \label{eq:Omega2}
 	\end{align}
 	with $(g,m)=(0,1)$ and $(0,2)$ were derived in \cite{Branahl:2020yru}. Evaluating \eqref{eq:Omega1} and \eqref{eq:Omega2} at the points $z_i=\varepsilon_i$ using \eqref{eq:R} and $R(\varepsilon_i)=e_i$ results in 
 	\begin{align}
 		\Omega_1^{(0)}(\varepsilon_q) &= \frac{\varepsilon_q - e_q}{\lambda} + \frac{1}{N} \sum_{n=1}^d r_n \left(\frac{1}{R'(\varepsilon_n)(\varepsilon_n - \varepsilon_q)} - \frac{1}{e_n-e_q} \right) \label{eq:Omega1_eps} \, , \\
 		\Omega_2^{(0)}(\varepsilon_q,\varepsilon_r) &= \frac{1}{R'(\varepsilon_q) R'(\varepsilon_r)} \left( \frac{1}{(\varepsilon_q-\varepsilon_r)^2} + \frac{1}{(\varepsilon_q+\varepsilon_r)^2} \right) \, . \label{eq:Omega2_eps} 
 	\end{align}
 	These expressions are still exact. To get a perturbative expansion in orders of $\lambda$, the $\varepsilon_i$'s and $R'(\varepsilon_i)$'s need to be expanded into a power series. Differentiating \eqref{eq:R} and inserting $\varepsilon_q$ gives
 	\begin{equation}\label{eq:Rs_exact}
 		R'(\varepsilon_q) = 1 + \frac{\lambda}{N} \sum_{n=1}^d \frac{r_n}{R'(\varepsilon_n)(\varepsilon_q + \varepsilon_n)^2} \, ,
 	\end{equation}
 	which can be used to calculate the expansion of $R'(\varepsilon_q)$ iteratively by inserting \eqref{eq:Rs_exact} into itself. To do this, expansions of the $\varepsilon_i$'s need to be iteratively calculated as well, which is done using \eqref{eq:eps_exact} that in turn depends on $R'(\varepsilon_n)$. With \eqref{eq:eps_exact} and \eqref{eq:Rs_exact} the expansions can be calculated up to arbitrary orders in $\lambda$. Inserting them into \eqref{eq:Omega1_eps} and \eqref{eq:Omega2_eps} results in perturbative expansions, which are equivalent to the expansions of $\Omega^{(0)}_{q}$ and $\Omega^{(0)}_{q,r}$ in every order when setting all multiplicities $r_i = 1$. 

\section{Vacuum graphs of the free energy}\label{ch:vacuumGraphs}

The first approach for expanding $\Omega_m^{(0)}$ into a sum of ribbon graphs utilizes \eqref{eq:generalDef_OmegaF}, where the boundary creation operator is repeatedly applied to the free energy $\mathcal{F}^{(0)}$ of genus $g=0$, which was given in \cite{Branahl:2021uxs} (cited in example \ref{exmp:freeEnergy}) up to order $\lambda^2$. In this section we expand upon those results by explicitly determining all graphs that contribute to $\mathcal{F}^{(g)}$ for $g\in\{0,1,2,3\}$ up to fifth order in $\lambda$. Combined with further information on each graph, like its genus $g$ or the order of its automorphism group, we can search this catalog of 5660 graphs for a subset with specific properties and can translate the catalog entries of a subset of graphs into the corresponding weights, thus obtaining the initial data for calculating the expansions of $\Omega_m^{(g)}$ using \eqref{eq:generalDef_OmegaF}.
 
 A significant inspiration for approaching the problem of determining vacuum ribbon graphs using the encoding introduced below was a paper by Coquereaux and Zuber \cite{Zuber2016}. They were interested in the subset of vacuum ribbon graphs that are embeddings of a single closed curve and thus related to knots.

\subsection{Encoding as a permutation pair}\label{ch:Encoding}

A ribbon graph has a unique Riemann surface in which it is embedded and thus contains additional information in comparison to the usual notion of a graph $(V,E)$, whose complete information is contained in a set $V$ of vertices and a set $E$ of edges, which are pairs of vertices.

In contrast, the basic building block of ribbon graphs is a set of vertices, that each have an arbitrary number of \emph{half edges} attached to them. The half edges are then pairwise connected to give a ribbon graph. Being an embedded graph the sequence of half edges at a vertex in clockwise rotation is fixed. This information of the vertices is encoded as a permutation $\sigma$ in the following way. One consecutively numbers all half edges in an arbitrary way. The cycles of $\sigma$ are then given by the clockwise sequence of half edges at each vertex.

\begin{figure}[H]
	\centering
	\def\svgwidth{0.7\columnwidth}
	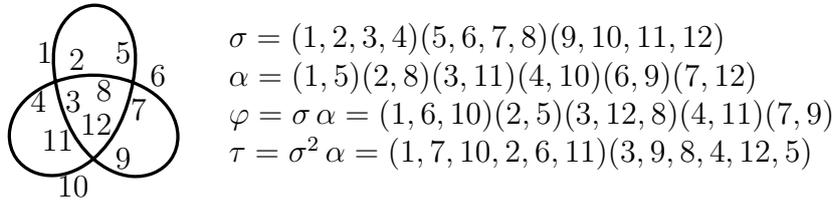
	\caption{Example of the encoding of a graph. $\sigma$ encodes the vertices, $\alpha$ the edges, $\varphi$ the faces and $\tau$ the sequence of labels along one side of a component.}
	\label{fig:EncodingExample.pdf_tex}
\end{figure}

A second permutation $\alpha$ encodes the pairing of half edges as cycles of length 2. The permutation pair $(\alpha,\sigma)$ then contains the whole information of the ribbon graph \cite{Bessis_1980}. An example is given in figure \ref{fig:EncodingExample.pdf_tex}. 

The permutation pair of a graph depends on the choice of the numbering of half edges, so it is not unique. Checking wether two distinct permutation pairs encode the same graph is a computational expensive calculation and the main bottleneck when going to high orders in $\lambda$. This will be briefly discussed in appendix \ref{ch:A_remarksImplementation}.

This concept can also be generalized to encode \emph{dessins d'enfants}, which have two sets of vertices (black and white). The sequence of edges incident to the white and black vertices is given by the cycles of $\alpha$ and $\sigma$ respectively. Ribbon graphs are closely related to a special type, called \emph{clean} dessins d'enfants, where all white vertices are two-valent (refer to chapter $4.1$ of \cite{IntroRiemannSurfaces}). In this picture half edges are considered to be edges.

The complete set of possible pairs of permutations $\alpha$ and $\sigma$ contains encodings of disconnected graphs and a lot of encodings that represent the same graph. By choosing the following conventions the size of the set is significantly reduced. 
\begin{convention}[Vacuum ribbon graphs]\label{conv:1}
	For a vacuum ribbon graph with $v$ four-valent vertices
	\begin{itemize}
		\item fix permutation $\sigma$, the encoding of the vertices, to $\sigma=(1,2,3,4)\cdots(4v-3,4v-2,4v-1,4v)$
		\item partially fix $\alpha$, the encoding of the edges, by pairing $v-1$ of the $2v$ edges according to the pattern $\alpha=\underbrace{(1,5)(5+j_1,9)\cdots(4v-7+j_{v-2},4v-3)}_{v-1\text{ 2-cycles}}\underbrace{\cdots}_{v+1 \text{ 2-cycles}}$ where $j_i\in\{1,2,3\} \, \forall i=1,\dots,v-2$.
	\end{itemize}
\end{convention}
Applying the first restriction results in a set of $(4v-1)!!$ permutation pairs \cite{Zuber2016} that includes pairs associated with disconnected graphs. These are removed with the second restriction which significantly reduces the size of the set of valid permutation pairs to $(2v+1)!!\,3^{v-2}$ pairs for $v>1$. 

The last step in obtaining a set of a minimum number of permutation pairs encoding all distinct vacuum ribbon graphs up to some order in $\lambda$ is to filter duplicate pairs that encode the same graph. Minding the convention, two permutation pairs encode the same graph if we can transform one into the other through a sequence of rotations of the half edge labels belonging to the same vertex and swaps of the labels between two vertices. These permutations generate a group
\begin{align*}
	G = &\langle (1,2,3,4),\dots,(4v-3,4v-2,4v-1,4v), \\
	&\quad (1,5)(2,6)(3,7)(4,8),\dots,(1,4v-3)(2,4v-2)(3,4v-1)(4,4v)\rangle,
\end{align*} 
here it is assumed that $v>1$. If two different permutations $\alpha_1,\alpha_2$ ($\sigma=\sigma_1=\sigma_2$ is fixed) belong to the same orbit under the action of $G$, then they encode the same graph. For higher orders in $\lambda$ and thus growing $v$ the order $|G|=4^v v!$ of the group is growing fast, so calculating the orbit of a permutation $\alpha$ becomes resource intensive. This is the motivation for filtering the initial set as much as possible.
 
The various properties of a graph that are calculated from a permutation pair $(\alpha,\sigma)$ will be discussed in the following.

\paragraph{Genus}

\begin{figure}
	\centering
	\def\svgwidth{0.4\columnwidth}
	%% Creator: Inkscape 1.1.2 (b8e25be8, 2022-02-05), www.inkscape.org
%% Accompanies image file 'SigmaAlpha.pdf' (pdf, eps, ps)
\begingroup%
  \makeatletter%
  \providecommand\color[2][]{%
    \errmessage{(Inkscape) Color is used for the text in Inkscape, but the package 'color.sty' is not loaded}%
    \renewcommand\color[2][]{}%
  }%
  \providecommand\transparent[1]{%
    \errmessage{(Inkscape) Transparency is used (non-zero) for the text in Inkscape, but the package 'transparent.sty' is not loaded}%
    \renewcommand\transparent[1]{}%
  }%
  \providecommand\rotatebox[2]{#2}%
  \newcommand*\fsize{\dimexpr\f@size pt\relax}%
  \newcommand*\lineheight[1]{\fontsize{\fsize}{#1\fsize}\selectfont}%
  \ifx\svgwidth\undefined%
    \setlength{\unitlength}{232.32902402bp}%
    \ifx\svgscale\undefined%
      \relax%
    \else%
      \setlength{\unitlength}{\unitlength * \real{\svgscale}}%
    \fi%
  \else%
    \setlength{\unitlength}{\svgwidth}%
  \fi%
  \global\let\svgwidth\undefined%
  \global\let\svgscale\undefined%
  \makeatother%
  \begin{picture}(1,0.46131864)%
    \lineheight{1}%
    \setlength\tabcolsep{0pt}%
    \put(0,0){\includegraphics[width=\unitlength,page=1]{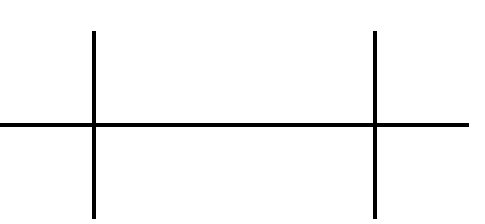}}%
    \put(0.62,0.135){\makebox(0,0)[t]{\lineheight{1.25}\smash{\begin{tabular}[t]{c}$k$\end{tabular}}}}%
    \put(0.7,0.32){\makebox(0,0)[t]{\lineheight{1.25}\smash{\begin{tabular}[t]{c}$k+1$\end{tabular}}}}%
    \put(0.92,0.23){\makebox(0,0)[t]{\lineheight{1.25}\smash{\begin{tabular}[t]{c}$k+2$\end{tabular}}}}%
    \put(0.32,0.25){\makebox(0,0)[t]{\lineheight{1.25}\smash{\begin{tabular}[t]{c}$j$\end{tabular}}}}%
    \put(0,0){\includegraphics[width=\unitlength,page=2]{SigmaAlpha.pdf}}%
    \put(0.39,0.03){\makebox(0,0)[t]{\lineheight{1.25}\smash{\begin{tabular}[t]{c}$\alpha$\end{tabular}}}}%
    \put(0.51,0.26720756){\makebox(0,0)[t]{\lineheight{1.25}\smash{\begin{tabular}[t]{c}$\sigma$\end{tabular}}}}%
    \put(0.90389038,0.435){\makebox(0,0)[t]{\lineheight{1.25}\smash{\begin{tabular}[t]{c}$\sigma$\end{tabular}}}}%
  \end{picture}%
\endgroup%

	\caption{Illustration of the face permutation $\varphi(j)=\sigma\circ\alpha(j)=k+1$ and permutation $\tau(j)=\sigma^2\circ\alpha(j)=k+2$, which goes along one side of a component (the horizontal line in this picture).}
	\label{fig:SigmaAlpha.pdf_tex}
\end{figure}

According to \eqref{eq:Euler} the genus $g$ of a vacuum ribbon graph $(n=0,b=0)$ with $v$ four-valent vertices $(r=2v)$ is $g=(v-l)/2+1$, where $l$ is the number of loops or in other words the number of faces. The permutation of face cycles of a pair $(\alpha,\sigma)$ is given by $\varphi=\sigma \alpha$ (see figure \ref{fig:EncodingExample.pdf_tex} for an example). An informal proof due to \cite{LandoZvonkin} is shown in figure \ref{fig:SigmaAlpha.pdf_tex}. The number of cycles of $\varphi$ (including one-cycles) is equivalent to $l$, so $g$ can be easily calculated.

\paragraph{Component number}

A vacuum ribbon graph is either a single closed curve, called \emph{component}, or made up of multiple overlapping closed curves. The number $c$ of components of a graph is encoded in a permutation $\tau=\sigma^2 \alpha$, whose cycles are the sequence of labels along a component. See figure \ref{fig:EncodingExample.pdf_tex} for an example and figure \ref{fig:SigmaAlpha.pdf_tex} for a general illustration of $\tau$. For each component $\tau$ has two cycles (one for the sequence of labels along the inner and one along the outer side of the circle), so $c$ is equivalent to half the number of cycles of $\tau$. 

\paragraph{Order of the automorphism group}

The automorphism group $\text{Aut}(\Gamma)$ of a graph $\Gamma$ contains the set of permutations $\gamma \in G$ which commute 
\begin{equation*}
	\alpha = \gamma \alpha \gamma^{-1}, \qquad \sigma = \gamma \sigma \gamma^{-1}
\end{equation*}
with the permutation pair $(\alpha,\sigma)$ corresponding to $\Gamma$ (see definition $2.3$ in \cite{Borot_2020}). Its order $|\text{Aut}(\Gamma)|$ is thus the number of automorphisms $\gamma$. This group is only non-trivial for vacuum ribbon graphs, which have no boundaries (cf.\ proposition $1.1.1$ in \cite{Eynard:2016yaa}). 

\paragraph{Testing bicolorability}

A vacuum ribbon graph with four-valent vertices is called \emph{bicolorable} if one can color its faces with two colors in such a way that at every vertex adjacent faces are colored differently and faces diagonally opposite of each other are colored the same \cite{Zuber2016}. This notion is closely related to bipartite graphs. A graph $(V,E)$ is \emph{bipartite} if its set of vertices $V$ can be partitioned into two sets $V_1, V_2$ such that no edge in $E$ connects two vertices of $V_1$ or $V_2$ \cite{Mohar2001GraphsOS}. If a ribbon graph is bicolorable then its dual ribbon graph, obtained by swapping vertices and faces, is bipartite \cite{ZinnJustinZuber}. Every planar vacuum ribbon graph with four-valent vertices is trivially bicolorable. In general, this is not the case for graphs with $g>0$. For example, the two faces adjacent to an edge can be the same face for non-planar graphs.

The method for testing this property is not very sophisticated, but a simple brute force calculation. The $l$ faces of a graph can be colored in $2^{l-1}$ ways using two colors (note that we fix the color of one face). If the strands of all ribbons (two-cycles of $\alpha$) for at least one of these options are not of the same color, then the graph is bicolorable.

\subsection{The catalog} 

The final catalog contains the encoding for all vacuum ribbon graphs up to fifth order in $\lambda$ with their respective genus, number of components, order of the automorphism group and wether the graph is bicolorable. Some examples from this catalog are given in appendix \ref{ch:B_examplesCatalog} with drawings of the corresponding graphs.

\begin{table}
    \begin{subtable}[h]{0.45\textwidth}
        \centering
        \begin{tabular}{|cc|ccccc|}
		\hline
		& & \multicolumn{5}{c|}{$m$} \\
		& & 1 & 2 & 3 & 4 & 5 \\
		\hline
		\multirow{4}{*}{$g$} & 0 & 1 & 3 & 7 & 33 & 156 \\
		& 1 & 1 & 4 & 23 & 185 & 1647 \\
		& 2 & & & 6 & 147 & 2937 \\
		& 3 & & & & & 510 \\
		\hline
	\end{tabular}
       \caption{Numbers of graphs.}
       \label{tab:vacuumGraphs}
    \end{subtable}
    \hfill
    \begin{subtable}[h]{0.45\textwidth}
        \centering
        \begin{tabular}{|cc|ccccc|}
		\hline
		& & \multicolumn{5}{c|}{$m$} \\
		& & 1 & 2 & 3 & 4 & 5 \\
		\hline
		\multirow{4}{*}{$g$} & 0 & $\frac{1}{2}$ & $\frac{9}{8}$ & $\frac{9}{2}$ & $\frac{189}{8}$ & $\frac{729}{5}$ \\
		& 1 & $\frac{1}{4}$ & $\frac{15}{8}$ & $\frac{33}{2}$ & $\frac{2511}{16}$ & $\frac{15633}{10}$\\
		& 2 &  & & $\frac{15}{4}$ & $\frac{2007}{16}$ & $\frac{28323}{10}$ \\
		& 3 &  &  &  &  & $\frac{945}{2}$ \\
		\hline
		\end{tabular}
        \caption{Sums of reciprocals.}
        \label{tab:AutVacuumGraphs}
     \end{subtable}
     \caption{Numbers of graphs and the sums of reciprocals of the order of their automorphism groups for the sets $\mathfrak{G}^{g,v}_\emptyset$.}
\end{table}

Table \ref{tab:vacuumGraphs} contains the total number of graphs for each order and genus. The first and second row correspond to the OEIS sequences A054935 and A292971 respectively, which give the numbers of planar and toroidal 4-regular (having only four-valent vertices) maps on an oriented surface. Of special interest is the
\begin{equation*}
	\text{OEIS sequence A}292206: \qquad 2, 7, 36, 365, 5\,250, 103\,801, 2\,492\,164,\dots \, ,
\end{equation*}
which are the numbers of \emph{unrooted} (no boundary) \emph{unlabeled connected four-regular maps on a compact closed oriented surface with $n$ vertices}. These are the totals of the columns of table \ref{tab:vacuumGraphs}. 

By considering the subset of vacuum ribbon graphs that have a single component $(c=1)$, one obtains the result for the number of \emph{unoriented immersions of a circle in an oriented surface of genus $g$} in table $8$ of \cite{Zuber2016} (UO rows in their notation) up to fifth order (see also OEIS sequence A260848). Furthermore, the numbers of bicolorable graphs, which are listed by the authors in table $9$ (UOb rows) of \cite{Zuber2016}, are equivalent to the catalog.

An additional crosscheck for the validity of the catalog and the calculated orders of automorphism groups is the sum of their reciprocals $1/|\text{Aut}(\Gamma_0)|$ of all graphs $\Gamma_0 \in \mathfrak{G}^{g,v}_\emptyset$ for a specific genus and order, which are listed in table \ref{tab:AutVacuumGraphs}. These can either be obtained directly from the catalog or from the free energy $\mathcal{F}^{(g)}$ in \eqref{eq:FreeEnergyGeneral} by reducing the number of pairwise different eigenvalues $(e_1,\dots,e_d)$ of the weights $\varpi(\Gamma_0)$ to a single eigenvalue $e$, so $d=1$. It was mentioned in \cite{Branahl:2021uxs, branahl2021genus} that in this limit the free energy of the QKM is equivalent to the free energy of quadrangulations in the Hermitian 1-matrix model.

\begin{proposition}[chapter 3.6.1, \cite{Eynard:2016yaa}]
The free energy of the Hermitian 1-matrix model with quartic potential and $g=0,1$ is given by
\begin{align*}
	&F_0 = t^2 \sum_{n=1}^{\infty} 3^n \frac{(2n-1)!}{n!(n+2)!} (tt_4)^n \, ,
	&F_1 = \frac{1}{24} \sum_{n=1}^{\infty} \frac{3^n}{n}\bigg(2^{2n}-\frac{(2n)!}{n!n!} \bigg) (tt_4)^n \, .
\end{align*} 
\end{proposition}
The rational coefficients of these formal series are equivalent to the coefficients of the special case with $d=1$ in the QKM 
\begin{align*}
	\mathcal{F}^{(0)}\big|_{d=1} &=\frac{-\lambda}{2(2E)^2}+\frac{9(-\lambda)^2}{8(2E)^4} +\frac{9(-\lambda)^3}{2(2E)^6} +\frac{189(-\lambda)^4}{8(2E)^8} +\frac{729(-\lambda)^5}{5(2E)^{10}} +\cdots \, ,\\
	\mathcal{F}^{(1)}\big|_{d=1} &=\frac{-\lambda}{4(2E)^2}+\frac{15(-\lambda)^2}{8(2E)^4} +\frac{33(-\lambda)^3}{2(2E)^6} +\frac{2511(-\lambda)^4}{16(2E)^8} +\frac{15633(-\lambda)^5}{10(2E)^{10}} +\cdots \, .\\
\end{align*}
These coefficients are listed in table \ref{tab:AutVacuumGraphs}. The expressions $F_g$ with $g=2,3$ are also calculated in \cite{Eynard:2016yaa} and their coefficients coincide with the rational numbers in table \ref{tab:AutVacuumGraphs} as well. 

\subsection{Graphs with boundaries encoded}\label{ch:EncodingBoundary}

It is natural to generalize the encoding of vacuum ribbon graphs in the previous sections to ribbon graphs, which have boundaries (see figure \ref{fig:2point_g1_lam2} for a depiction with strands instead of half edges) and thus are part of the more general set $\mathfrak{G}^{g,v}_{|p_1^1...p_{n_1}^1|...|p_1^b...p_{n_b}^b|}$. 

\begin{convention}[Ribbon graphs with boundaries]
	In addition to the chosen convention for vacuum ribbon graphs we adopt the following convention for the case of ribbon graphs with $v$ four-valent and $n$ one-valent vertices due to one or multiple boundaries: 
	\begin{itemize}
		\item add $n$ one-cycles to the vertices permutation representing one-valent vertices, so\\ $\sigma =(1,2,3,4)\cdots(4v-3,4v-2,4v-1,4v)(4v+1)\cdots(4v+n)$
		\item add $n/2$ two-cycles to the permutation $\alpha$ to encode the additional half edges.
	\end{itemize}
\end{convention}
This encoding provides a second approach for obtaining the numbers of ribbon graphs with boundaries whose weights contribute to the perturbative expansion of the $\Omega^{(g)}_m$ in \eqref{eq:generalDef_OmegaF}. Instead of letting the BCO act as a partial derivative on a sum of weights, we can define it in the language of encodings of ribbon graphs. We discuss this in the following section.

\section{Perturbative expansion}\label{ch:PerturbativeExpansion}

Before the discussion of the perturbative expansions of $\Omega^{(g)}_m$ for various $g$ and $m$ up to fifth order in subsections \ref{ch:CountingGraphs} and \ref{ch:Comparison}, we give in the following subsection a visual interpretation of the action of the BCO both in the language of weights and encodings with permutations.

\subsection{Boundary creation operator}
The boundary creation operator $\hat{T}_q=- N \frac{\partial}{\partial E_q}$ was introduced as part of the recursive definition \eqref{eq:generalDef_OmegaF} of $\Omega^{(g)}_{q_1,\dots,q_m}$, where it acts on weights $\varpi(\Gamma)$. We demand, that the indices $q_1,\dots,q_m$ are pairwise different. This condition ensures that $\hat{T}$ can only act on free indices. For every one of those indices the partial derivative eliminates all terms of the sum except the one with matching index. The boundary creation operator thus fixes a summation index and raises its exponent in the denominator by one. This can be translated into pictures. Some remarks concerning notation: We omit all factors of $N$, the order of the automorphism group and all summation signs.
The index $q$ will be associated with a fixed index and the indices $j,k$ and $l$ are free indices. To distinguish them in the drawing, each index will be associated with a line style
\newcommand{\lwLegende}{1.5}
\begin{alignat*}{4}
	&j\quad &\tikz\draw[n, line width=\lwLegende] (0,0)--(1,0);\,, &\qquad k\quad &\tikz\draw[k, line width=\lwLegende] (0,0)--(1,0);\,, &\qquad l\quad &\tikz\draw[l, line width=\lwLegende] (0,0)--(1,0);\,, &\qquad q\quad \tikz\draw[q, line width=\lwLegende] (0,0)--(1,0);\,.
\end{alignat*}
As a small example we apply $\hat{T}_q$ to the graphs in zeroth and first order of $\mathcal{F}^{(0)}$.

\begin{tikzpicture}[scale=1.85, line width=\lwMacro, font=\normalsize, miter limit=\miterLimit, baseline]

\renewcommand{\posUp}{-0.3}
\renewcommand{\posDown}{0.225}

\renewcommand{\vHeight}{-0.4}	
	\ZeroZero{(\posZero,\vHeight)}{n}{k}
	\draw[cut] (0.5,\vHeight+0.1) -- (0.5,\vHeight+0.5) node[above] {\RNum{1}};
	\node[right=\eqGraphs, scale=\mathScale] at (\posZero,\vHeight) {$ \displaystyle \sim \repNK \log(E_j {+} E_k) $};
	\node[red,left] at (\resOne-\numGap,\vHeight) {\RNum{1}};
	\coordinate (down) at (\resOne,\vHeight-\BraceWidth);
	\coordinate (up) at (\resOne,\vHeight+\BraceWidth);
	\draw[klammer,line width=\lwBrackets] (down) node[above right] {cut $j \to$} -- (up) node[below right] {cut $k \to$};
	\scoped[scale=\smallScale] \OneZero{($(up)+(\posMiniGraph,\posUp)$)}{q}{n};
	\scoped[scale=\smallScale] \OneZero{($(down)+(\posMiniGraph,\posDown)$)}{k}{q};
	\coordinate (down) at (\resOne+\braceGap,\vHeight-\BraceWidth);
	\coordinate (up) at (\resOne+\braceGap,\vHeight+\BraceWidth);
	\draw[klammer,line width=\lwBrackets] (up) -- (down) node[midway,right=\eqBrace,scale=\mathScale] {$ \displaystyle \sim 2 \times \repN \frac{1}{E_q{+}E_j} $};
	
\renewcommand{\posUp}{-0.55}
\renewcommand{\posDown}{0}	
	
\renewcommand{\vHeight}{-1.6}
	\ZeroOne{(\posZero,\vHeight)}{n}{k}{l}
	\draw[cut] (0.25,\vHeight) -- (0.25,\vHeight+0.4) node[above] {\RNum{1}};
	\draw[cut] (0.75,\vHeight) -- (0.75,\vHeight+0.4) node[above] {\RNum{2}};
	\node[right=\eqGraphs, scale=\mathScale] at (\posZero,\vHeight) {$ \displaystyle \sim \repNKL \frac{1}{(E_j{+}E_k)(E_j{+}E_l)} $};
	
\renewcommand{\vHeight}{-2.8}
	\node[red,left] at (\resZero-\numGap,\vHeight) {\RNum{1}};
	\coordinate (down) at (\resZero,\vHeight-\BraceWidth);
	\coordinate (up) at (\resZero,\vHeight+\BraceWidth);
	\draw[klammer,line width=\lwBrackets] (down) node[above right] {cut $j \to$} -- (up) node[below right] {cut $k \to$};
	\scoped[scale=\smallScale] \OneOne{($(up)+(\posMiniGraph,\posUp)$)}{q}{n}{l};
	\scoped[scale=\smallScale] \OneOne{($(down)+(\posMiniGraph,\posDown)$)}{k}{q}{l};
	\coordinate (down) at (\resZero+\braceGap,\vHeight-\BraceWidth);
	\coordinate (up) at (\resZero+\braceGap,\vHeight+\BraceWidth);
	\draw[klammer,line width=\lwBrackets] (up) -- (down);
	
\renewcommand{\vHeight}{-2.8}
	\node[red,left] at (\resOne-\numGap,\vHeight) {\RNum{2}};
	\coordinate (down) at (\resOne,\vHeight-\BraceWidth);
	\coordinate (up) at (\resOne,\vHeight+\BraceWidth);
	\draw[klammer,line width=\lwBrackets] (down) node[above right] {cut $j \to$} -- (up) node[below right] {cut $l \to$};
	\scoped[scale=\smallScale] \OneOne{($(up)+(\posMiniGraph,\posUp)$)}{q}{n}{k};
	\scoped[scale=\smallScale] \OneOne{($(down)+(\posMiniGraph,\posDown)$)}{l}{q}{k};
	\coordinate (down) at (\resOne+\braceGap,\vHeight-\BraceWidth);
	\coordinate (up) at (\resOne+\braceGap,\vHeight+\BraceWidth);
	\draw[klammer,line width=\lwBrackets] (up) -- (down);
	
\renewcommand{\vHeight}{-3.8}
	\coordinate (down) at (\resZero+\braceGap,\vHeight-\BraceWidth);
	\coordinate (up) at (\resZero+\braceGap,\vHeight+\BraceWidth);
	\node[right=\eqGraphs,scale=\mathScale] at (\posZero,\vHeight) {$ \displaystyle \sim 2 \times \repNK \frac{1}{(E_q{+}E_j)^2} \bigg(\frac{1}{E_j{+}E_k} + \frac{1}{E_q{+}E_k} \bigg) $};	
\end{tikzpicture}\\
In zeroth order there is only one ribbon that $\hat{T}_q$ can cut in half, but the operator itself can act either on the outer strand with index $j$ or the inner with $k$. This results in two graphs: two ribbons whose ends are one-valent vertices and which both have a strand with fixed index $q$ and a strand with a free index (either $j$ or $k$). We can rename the index $k\to j$ and thus the same graph is produced twice. This gives a new interpretation of the factor of the reciprocal of $|\text{Aut}(\Gamma_0)|$. It is necessary to prevent an overcounting of equivalent ribbon graphs that are obtained by applying the BCO.
The second order in $\lambda$ is analogous to the zeroth and first. Its illustrations are part of appendix \ref{ch:appendixPictures}, where we also apply $\hat{T}_r$ to the graphs of $\Omega^{(0)}_q$, thus obtaining the weights of $\Omega^{(0)}_{q,r}$.
\bigskip

Instead of translating a ribbon graph into a weight and acting with the BCO as a partial derivative, one can also define a BCO which acts on an encoding of a ribbon graph. With this alternative no information about the underlying ribbon graphs is lost. 

To define such an operator we need to keep track of the half edges on which it can act. This additional information is contained in a set $\beta$ of sets, so a ribbon graph with boundaries is encoded as a triple $(\alpha,\sigma,\beta)$. As an example we again consider the vacuum ribbon graph at first order in the picture above. It can be encoded as a pair $(\alpha,\sigma)$ with $\sigma=(1,2,3,4)$ and $\alpha=(1,2)(3,4)$. It has three faces $\varphi=(1,3)(2)(4)$, or three loops $l=3$, and thus its weight has three free indices (compare with the second term in \eqref{eq:free_energy}). The BCO can act on half edges/strands that correspond to a free index, so initially $\beta=\{\{1,3\},\{2\}\{4\}\}$. By applying it to a half edge once we get two additional one-valent vertices, $(5)$ and $(6)$, an extra edge and the free index associated with the half edge is fixed, thus the set in $\beta$, that contains said half edge, is removed. We obtain four triples
\begin{align*}
	&((1,2,3,4)(5)(6),(5,2)(3,4)(1,6),\{\{2\}\{4,6\}\}) \\
	&((1,2,3,4)(5)(6),(1,5)(3,4)(2,6),\{\{1,3,6\}\{2\}\}) \\
	&((1,2,3,4)(5)(6),(1,2)(5,4)(3,6),\{\{2\}\{4,6\}\}) \\
	&((1,2,3,4)(5)(6),(1,2)(3,5)(4,6),\{\{1,3,6\}\{4\}\}) 
\end{align*}  
by separately acting on the half edges numbered $1,2,3$ and $4$. The first and third (second and fourth) encoding describe the same unlabeled ribbon graph. Again, we can remove these duplicates by only keeping a single encoding for each orbit under the action of $G$, as discussed in section \ref{ch:Encoding}, leaving us with encodings for the two distinct graphs depicted in the example above. The first/third (second/fourth) encoding corresponds to the second (first) term of the sum of weights in the picture above.

Some remarks about the underlying algorithm are necessary. It appends a new pair of half edges as a cycle $(4v+n-1,4v+n)$ to $\alpha$ ($(5,6)$ in our example) and then switches the labels $4v+n-1 \leftrightarrow i$, where $i$ is the label of the half edge on which the BCO acts. So by convention the half edge labelled $4v+n-1$ always corresponds to a fixed index. Conversely, one might assume that the half edge $4v+n$ always corresponds to a free index and thus always needs to be added to the corresponding set in $\beta$ (as is the case for our example), but this is generally not true and needs to be checked in each case individually.

\subsection{Counting graphs}\label{ch:CountingGraphs}

\paragraph{Number of graphs}

Table \ref{tab:NumberGraphsOmega} contains the total numbers of graphs of $\Omega^{(g)}_{q_1,\dots,q_m}$ for $0\leq m \leq 6$ and $g \geq 0$ up to order $\lambda^5$. There are also graphs whose weights contribute to the expansion of $\Omega^{(0)}_7$, but due to insufficient computational resources (mainly storage and RAM) it was not possible to calculate their number. 

\begin{table}[b]
	\centering
	\begin{tabular}{|c|c|ccccccc|}
		\hline
		$g$ & $v$ & \multicolumn{7}{c|}{$m$} \\
		& & 0 & 1 & 2 & 3 & 4 & 5 & 6 \\
		\hline
		\multirow{6}{*}{0} & 0 &  & 1 & 1 & & & & \\
		& 1 & 1 & 2 & 7 & 12 & & & \\
		& 2 & 3 & 9 & 58 & 240 & 486 & & \\
		& 3 & 7 & 54 & 522 & 3\,628 & 16\,272 & 35\,520 & \\
		& 4 & 33 & 378 & 4\,941 & 49\,464 & 358\,056 & 1\,675\,680 & 3\,821\,760 \\
		& 5 & 156 & 2916 & 48\,411 & 641\,196 & 6\,542\,208 & 48\,475\,968 & 232\,983\,360 \\
		\hline 
		\multirow{5}{*}{1} & 1 & 1 & 1 & & & & & \\
		& 2 & 4 & 15 & 44 & & & & \\
		& 3 & 23 & 198 & 1\,257 & 3\,512 & & & \\
		& 4 & 185 & 2\,511 & 25\,065 & 148\,968 & 410\,364 & & \\
		& 5 & 1\,647 & 31\,266 & 429\,381 & 3\,998\,052 & 23\,101\,632 & 63\,205\,824 & \\
		\hline 
		\multirow{3}{*}{2} & 3 & 6 & 45 & & & & & \\
		& 4 & 147 & 2\,007 & 9\,360 & & & & \\
		& 5 & 2937 & 56\,646 & 518\,598 & 1\,964\,520 & & & \\
		\hline 
		3 & 5 & 510 & 9\,450 & & & & & \\
		\hline
	\end{tabular}
	\caption{Numbers of individual graphs contributing to $\Omega^{(g)}_{q_1,\dots,q_m}$ for $0 \leq m\leq 6$ and $g \geq 0$.}
	\label{tab:NumberGraphsOmega}
\end{table}

 The second, third and fourth column ($m\in\{1,2,3\}$) in table \ref{tab:NumberGraphsOmega} are already known and can be calculated with a method introduced in \cite{Branahl:2021uxs}. Instead of determining the perturbative expansions $\Omega^{(0)}_{q_1,\dots,q_m}$ with \eqref{eq:generalDef_OmegaF}, one can express them using correlation functions $G^{(0)}_{\dots}$. Starting from the general relation for $\Omega^{(g)}_q$ in \eqref{eq:generalDef_OmegaF}, the authors evaluated the application of the boundary creation operator $\hat{T}$ to a general $G^{(g)}_{\dots}$ using an \emph{equation of motion} \cite{Schurmann:2021mzu}. As a result they obtained general expressions for $\Omega^{(0)}_{q_1,\dots,q_m}$ with arbitrary $g$ and $m\in \{1,2,3\}$. For example, $\Omega^{(0)}_{q,r}$ with $g=0$ is given by
\begin{align*}
	\Omega_{q,r}^{(0)} = \frac{1}{(E_q-E_r)^2} + G_{|qr|}^{(0)} G_{|qr|}^{(0)} + \frac{1}{N^2} \sum_{n,k=1}^N G_{|qn|rk|}^{(0)} + \frac{1}{N} \sum_{n=1}^N \left( G_{|qnqr|}^{(0)} + G_{|rnrq|}^{(0)} + G_{|qnrn|}^{(0)} \right) \, .
\end{align*}
The correlation functions can be explicitly calculated in the special case of $d=1$ (single energy eigenvalue $E_1=E$) with 
\begin{align*}
	G^{(0)}_{|k_1^1...k_{n_1}^1|...|k_1^b...k_{n_b}^b|} \Big|_{d=1} =\sum_{n=0}^{\infty} \frac{3^{b+n-2}(\#n_e-1)!}{n!(3l_h+b+n-2)!} \prod_{i=1}^b n_i \binom{\frac{3n_i}{2}}{\frac{n_i}{2}} \cdot \frac{(-\lambda)^{n+l_h+b-2}}{(2E)^{2(n+l_h+b-1)}}\;,
\end{align*}
a result, which is mentioned in \cite{Branahl:2021} and based on investigations by Bernardi and Fusy \cite{bernardi2017bijections}. Here $l_h = \frac{1}{2} \sum_i n_i$ is half the sum of all boundary lengths $n_i$ and $\#n_e = 3l_h +2b +2n -4$ is the total number of edges. By inserting these expansions into the expressions of $\Omega^{(0)}_{q_1,\dots,q_m}$ one obtains the numbers of graphs of the columns with $m\in\{1,2,3\}$. Furthermore, it was mentioned in \cite{Branahl:2021uxs} that the second column with $m=1$ and genus $g=0$ are the first numbers of rooted quadangulations of the sphere (c.f.\ chapter 3.1.7 in \cite{Eynard:2016yaa}, originally discovered by Tutte \cite{Tutte:1963??})
\begin{equation*}
	2\cdot 3^n\cdot \frac{(2n)!}{n!(n+2)!} = 2 \cdot 3^n \frac{C_n}{n+2} \,,
\end{equation*}
where $C_n$ are the \emph{Catalan numbers} and $n$ is the number of faces (in the terminology of ribbons graphs the number of vertices or the order in $\lambda$).

Note that according to \eqref{eq:generalDef_OmegaF} $\Omega^{(0)}_q$ only receives contributions from graphs of the $2$-point function $G^{(0)}_{|qk|}$. This agrees with the graphs of $\Omega^{(0)}_q$ illustrated in the previous subsection and appendix \ref{ch:appendixPictures}, which were obtained by cutting the vacuum graphs of figure \ref{fig:Lambda012}. They all belong to $G^{(0)}_{|qk|}$ (compare with example 3.5 in \cite{Branahl:2021uxs}).

The numbers of table \ref{tab:NumberGraphsOmega} were obtained from the sums of weights calculated with \eqref{eq:generalDef_OmegaF} by considering the case $d=1$. The same numbers are reproduced up to fourth order with the alternative implementation of the BCO, which acts on encodings instead of weights.

\paragraph{Number of weights} 

Table \ref{tab:NumberTermsOmega} displays the numbers of individual weights in the perturbative expansions of the $\Omega^{(g)}_m$. These are generally smaller than the numbers of graphs in table \ref{tab:NumberGraphsOmega} because the majority of weights $\varpi(\Gamma)$ does not uniquely determine a single graph. Instead, one can often associate many different graphs to a single weight as there are multiple ways of connecting the ribbons encoded by $\varpi(\Gamma)$. We do not know of an interesting interpretation of these numbers, but still list them here, because they were unobtainable with previous methods. 

\begin{table}
	\centering
	\begin{tabular}{|c|c|ccccccc|}
		\hline
		$g$ & $v$ & \multicolumn{7}{c|}{$m$} \\
		& & 0 & 1 & 2 & 3 & 4 & 5 & 6 \\
		\hline
		\multirow{6}{*}{0} & 0 &  & 1 & 1 & & & & \\
		& 1 & 1 & 2 & 5 & 6 & & & \\
		& 2 & 3 & 8 & 30 & 81 & 117 & & \\
		& 3 & 7 & 36 & 195 & 857 & 2\,652 & 4\,180 & \\
		& 4 & 29 & 188 & 1\,363 & 8\,465 & 41\,316 & 138\,335 & 235\,260 \\
		& 5 & 113 & 1086 & 10\,041 & 81\,770 & 555\,086 & 2\,912\,027 & 10\,397\,508 \\
		\hline 
		\multirow{5}{*}{1} & 1 & 1 & 1 & & & & & \\
		& 2 & 4 & 8 & 11 & & & & \\
		& 3 & 17 & 60 & 188 & 312 & & & \\
		& 4 & 88 & 502 & 2\,586 & 9\,470 & 17\,911 & & \\
		& 5 & 541 & 4\,447 & 32\,914 & 193\,728 & 789\,216 & 1\,633\,233 & \\ 
		\hline
		\multirow{3}{*}{2} & 3 & 1 & 1 & & & & & \\
		& 4 & 15 & 32 & 39 & & & & \\
		& 5 & 180 & 738 & 2\,126 & 3\,105 & & & \\
		\hline 
		3 & 5 & 1 & 1 & & & & & \\
		\hline
	\end{tabular}
	\caption{Numbers of individual weights contributing to $\Omega^{(g)}_{q_1,\dots,q_m}$ for $0 \leq m\leq 6$ and $g \geq 0$.}
	\label{tab:NumberTermsOmega}
\end{table}

\subsection{Comparison with expansion of exact BTR results}\label{ch:Comparison}
Now that the perturbative expansions in sums of weights of the $\Omega^{(g)}_m$ is known to fifth order, we can compare this result with the direct expansion of $\Omega^{(0)}_1$ and $\Omega^{(0)}_2$ according to section \ref{ch:TwoApproaches}. The difference of the perturbative expansion and expansion of the exact result vanishes in all cases up to fifth order, thus proving their equivalence.

We refer to the provided programs \cite{Zenodo} (appendix \ref{ch:A_remarksImplementation} gives a short overview) for the full expressions of both types of expansions. Alternatively, these are given explicitly in \cite{Branahl:2021uxs} up to first order in $\lambda$.

\section{Conclusion and outlook}\label{ch:conclusion}
We implemented closed expressions of $\Omega^{(0)}_1$ and $\Omega^{(0)}_2$ and verified the equivalence of their perturbative expansions to the sum of graph weights obtained with \eqref{eq:generalDef_OmegaF} up to order $\lambda^5$. Table \ref{tab:NumberGraphsOmega} lists the number of individual graphs and table \ref{tab:NumberTermsOmega} the number of distinct weights, which contribute to any of those $\Omega^{(g)}_m$. The corresponding expressions, i.e.\ the sum of weights (although not included in this paper) were also calculated. We obtained a catalog of encodings for all vacuum graphs up to fifth order and introduced a BCO, which can act on the $\alpha$ permutation of an encoding of a ribbon graph.

This is a first important step for the development of a program that automates the calculation of the perturbative expansion of the $\Omega^{(g)}_m$. There are two directions in which we can expand upon the current results.

It was briefly mentioned in section \ref{ch:BTR} that the meromorphic functions $\Omega^{(g)}_m$ are known for $(g,m) \in \{(0,2),(0,3),(0,4),(1,1)\}$, so there are at least three other functions, whose expansion can be calculated and compared to the results of perturbation theory of section \ref{ch:PerturbativeExpansion}. Furthermore, there is an explicit recursion formula for all $\Omega^{(0)}_m$, but with increasing $m$ the calculations are too cumbersome, so implementing this recursion in Mathematica would be certainly interesting. 

 We consider the implementation of a BCO applicable to permutation encodings of ribbon graphs in section \ref{ch:EncodingBoundary} as a first proof of concept, but in its current form the resulting dataset of ribbon graphs is not very useful (apart from determining total numbers of graphs, which can be done more efficiently by differentiating weights). Adding methods for determining various properties of these ribbon graphs, similar to the catalog, could provide a valuable tool in the perturbative analysis of the QKM. For example, the numbers of rooted maps for arbitrary genus are a known result (OEIS A238396). These correspond to the numbers of rooted bipartite quadrangulations \cite{branahl2021genus}. Cataloguing ribbon graphs with boundaries could be useful in understanding the relation between those and the bicolorable subset of all ribbon graphs, whose numbers are in the $m=1$ column of table \ref{tab:NumberGraphsOmega}.  

\newpage
\appendix

\section{Remarks on implementation}\label{ch:A_remarksImplementation}

All calculations were done with Mathematica \cite{Mathematica} using four different notebooks. These notebooks, a file containing the catalog of encodings of vacuum ribbon graphs and compressed files of the encodings and weights of the graphs contributing to $\Omega^{(g)}_m$ can be downloaded at \cite{Zenodo}. We give a short overview of the purpose of each notebook.

The notebook \texttt{Catalog.nb} calculates the catalog, so it determines a permutation pair $(\alpha,\sigma)$ for all vacuum graphs up to fifth order using the methods introduced in section \ref{ch:Encoding}. More precisely, two slightly different methods ($X$ and $X2$) are implemented. The latter has the advantage of utilizing the second restriction of convention \ref{conv:1}, thus reduces the number of permutations whose orbits under the action of $G$ have to be calculated. This is time consuming, so method $X2$ is significantly more efficient than method $X$. The notebook also provides a way of filtering the catalog according to the properties of the graphs. We present a small sample of the catalog in the next section of the appendix. It is uncertain if the sixth order with $103\,801$ graphs is obtainable with our current methods and limited computational resources. The main difficulty is clearly the fast growth of the order of $G$ ($|G|=4^n n!$ at $n$-th order in $\lambda$ \cite{Zuber2016}). 

The second notebook \texttt{CountingGraphs.nb} determines for every vacuum ribbon graph $\Gamma_0$ of the catalog the corresponding weight $\varpi(\Gamma_0)$ and explicitly calculates the graph expansion of $\Omega^{(g)}_m$ by applying the boundary creation operator for weights \eqref{eq:generalDef_OmegaF}, a partial derivative, to the sum of weights. One obtains perturbative expansions of the $\Omega_m^{(g)}$ up to fifth order in $\lambda$. Both the numbers of individual graphs and individual weights, given in table \ref{tab:NumberGraphsOmega} and table \ref{tab:NumberTermsOmega}, are calculated. 

As previously mentioned, some information is lost when translating a permutation pair of a vacuum ribbon graph to a weight. The notebook \texttt{CatalogBCO.nb} implements the boundary creation operator for triples $(\alpha,\sigma,\beta)$, as discussed in section \ref{ch:EncodingBoundary}, so all information is preserved. This approach is only sensible if one is interested in the actual structure of the graphs and their properties. 

The fourth notebook \texttt{CompareOmega.nb} imports the relevant perturbative expansions $\Omega^{(0)}_q$ and $\Omega^{(0)}_{q,r}$ that were obtained in \texttt{CountingGraphs.nb} and compares them to expansions of the exact expressions $\Omega^{(0)}_1$ and $\Omega^{(0)}_2$, calculated according to section \ref{ch:TwoApproaches}. This is done by considering their differences and replacing symbolic variables with integer values at a suitable moment in the calculation to improve performance. This is a valid method as calculations with rational numbers are exact in Mathematica, so no errors due to machine precision can occur.

\section{Some examples from the catalog}\label{ch:B_examplesCatalog}
Every vacuum graph $\Gamma_0$ with $v$ vertices, thus of order $v$, and genus $g$ has a catalog entry of the form $\{\alpha, v, g, c, |\text{Aut}(\Gamma_0)|, \theta\}$, with $\theta$ being a Boolean value which states wether the graph is bicolorable or not and the number of components $c$. The permutation $\sigma$, which encodes the sequence of incident half edges at each vertex, is by convention \ref{conv:1} the same for all graphs and thus not explicitly stated. As previously mentioned, the choice of $\alpha$, the permutation that encodes the edges, for some graph is not unique.

The following figures are some examples of subsets of the catalog. Every presented graph is drawn in the style of figure \ref{fig:Lambda012} of the introductory example \ref{exmp:freeEnergy} with its corresponding catalog entry. Graphs with $g>0$ cannot properly be projected on the plane without introducing additional crossing. Such unwanted vertices are called \emph{virtual} in analogy to virtual knot theory \cite{Kauffman_1999,ZinnJustinZuber} and are encircled red.

\begin{figure}[H]
	\centering
		\def\svgwidth{0.915\columnwidth}
		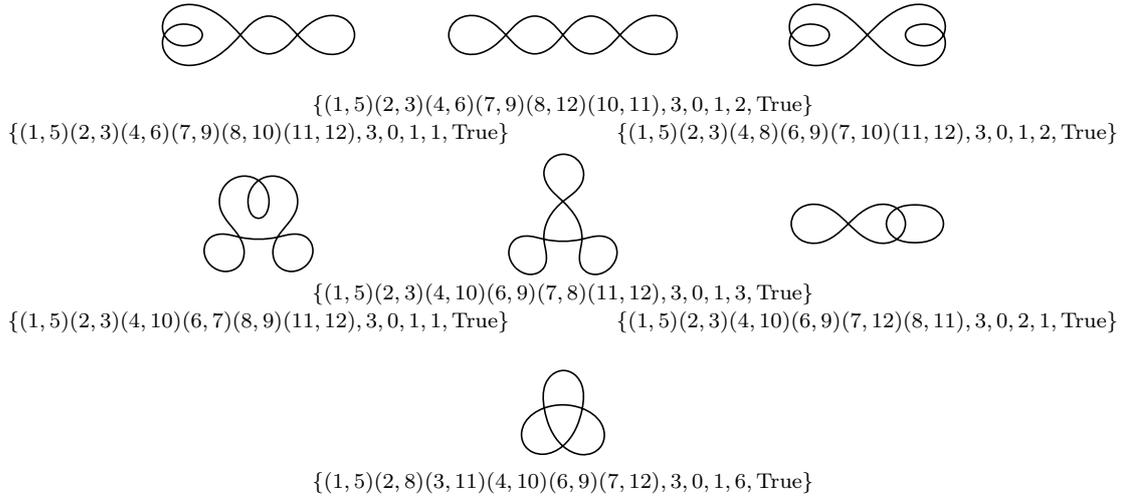
	\caption{All 7 vacuum ribbon graphs with three vertices and genus 0.}
\end{figure}

\begin{figure}[H]
	\centering
		\def\svgwidth{0.915\columnwidth}
		%% Creator: Inkscape 1.1.1 (c3084ef, 2021-09-22), www.inkscape.org
%% Accompanies image file 'Catalog23.pdf' (pdf, eps, ps)
\begingroup%
  \makeatletter%
  \providecommand\color[2][]{%
    \errmessage{(Inkscape) Color is used for the text in Inkscape, but the package 'color.sty' is not loaded}%
    \renewcommand\color[2][]{}%
  }%
  \providecommand\transparent[1]{%
    \errmessage{(Inkscape) Transparency is used (non-zero) for the text in Inkscape, but the package 'transparent.sty' is not loaded}%
    \renewcommand\transparent[1]{}%
  }%
  \providecommand\rotatebox[2]{#2}%
  \newcommand*\fsize{\dimexpr\f@size pt\relax}%
  \newcommand*\lineheight[1]{\fontsize{\fsize}{#1\fsize}\selectfont}%
  \ifx\svgwidth\undefined%
    \setlength{\unitlength}{797.58230063bp}%
    \ifx\svgscale\undefined%
      \relax%
    \else%
      \setlength{\unitlength}{\unitlength * \real{\svgscale}}%
    \fi%
  \else%
    \setlength{\unitlength}{\svgwidth}%
  \fi%
  \global\let\svgwidth\undefined%
  \global\let\svgscale\undefined%
  \makeatother%
  \begin{picture}(1,0.35381838)%
    \lineheight{1}%
    \setlength\tabcolsep{0pt}%
    \put(0.50052342,0.21888667){\makebox(0,0)[t]{\lineheight{1.25}\smash{\begin{tabular}[t]{c}\tiny$\{(1,5)(2,4)(3,7)(6,9)(8,11)(10,12),3,2,4,4,\text{False}\}$\end{tabular}}}}%
    \put(0.50052342,0.03081823){\makebox(0,0)[t]{\lineheight{1.25}\smash{\begin{tabular}[t]{c}\tiny$\{(1,5)(2,6)(3,10)(4,12)(7,9)(8,11),3,2,1,1,\text{False}\}$\end{tabular}}}}%
    \put(0.19961405,0.19067641){\makebox(0,0)[t]{\lineheight{1.25}\smash{\begin{tabular}[t]{c}\tiny$\{(1,5)(2,4)(3,6)(7,9)(8,11)(10,12),3,2,3,2,\text{False}\}$\end{tabular}}}}%
    \put(0.80143282,0.19067641){\makebox(0,0)[t]{\lineheight{1.25}\smash{\begin{tabular}[t]{c}\tiny$\{(1,5)(2,4)(3,10)(6,9)(7,11)(8,12),3,2,2,1,\text{False}\}$\end{tabular}}}}%
    \put(0.80143282,0.00260797){\makebox(0,0)[t]{\lineheight{1.25}\smash{\begin{tabular}[t]{c}\tiny$\{(1,5)(2,7)(3,10)(4,12)(6,9)(8,11),3,2,2,2,\text{False}\}$\end{tabular}}}}%
    \put(0.19961405,0.00260797){\makebox(0,0)[t]{\lineheight{1.25}\smash{\begin{tabular}[t]{c}\tiny$\{(1,5)(2,4)(3,10)(6,11)(7,12)(8,9),3,2,3,2,\text{False}\}$\end{tabular}}}}%
    \put(0,0){\includegraphics[width=\unitlength,page=1]{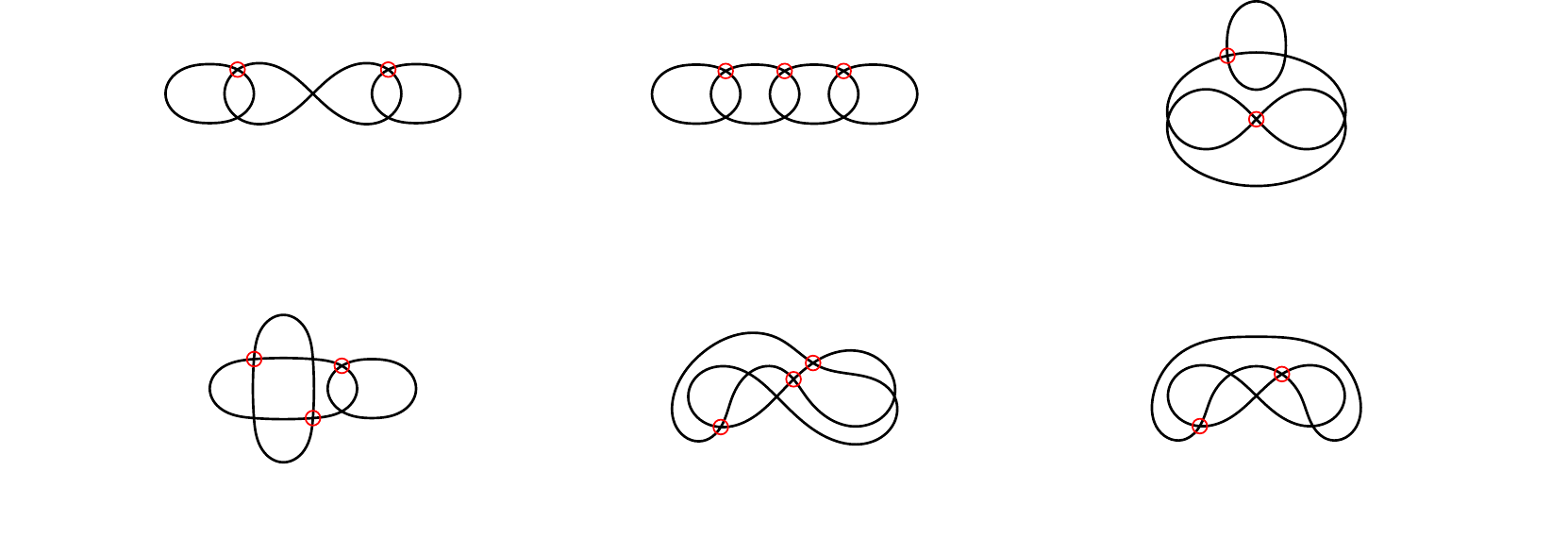}}%
  \end{picture}%
\endgroup%

		\caption{All 6 vacuum ribbon graphs with three vertices and genus 2.}
\end{figure}

\begin{figure}[H]
	\centering
		\def\svgwidth{0.915\columnwidth}
		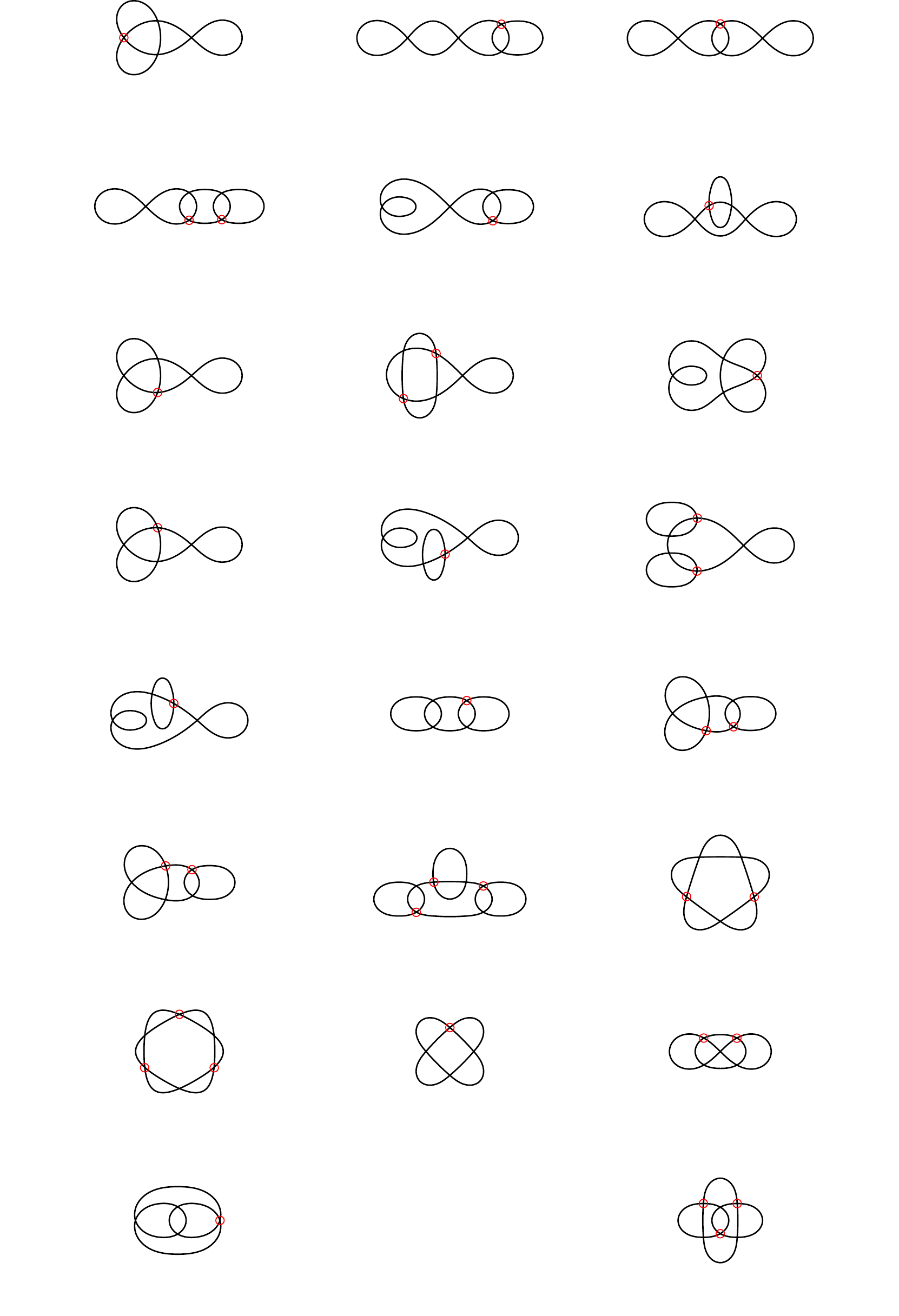
		\caption{All 23 vacuum ribbon graphs with three vertices and genus 1.}
\end{figure}

\begin{figure}[H]
	\centering
		\def\svgwidth{1.065\columnwidth}
		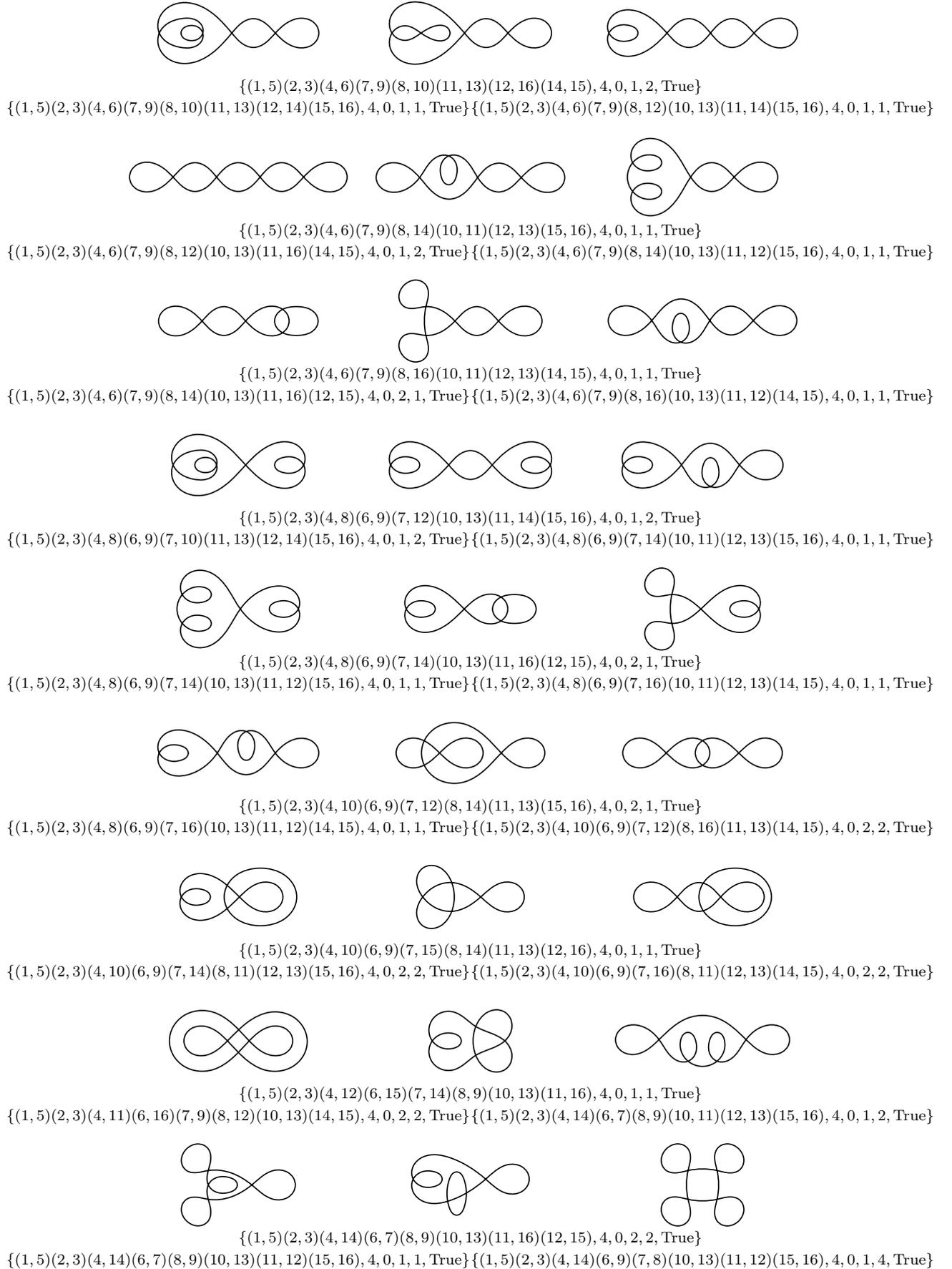
		\caption{All 33 vacuum ribbon graphs with four vertices and genus 0.}
		\label{fig:example41}
\end{figure}

\begin{figure}[H]
	\centering
		\def\svgwidth{1.065\columnwidth}
		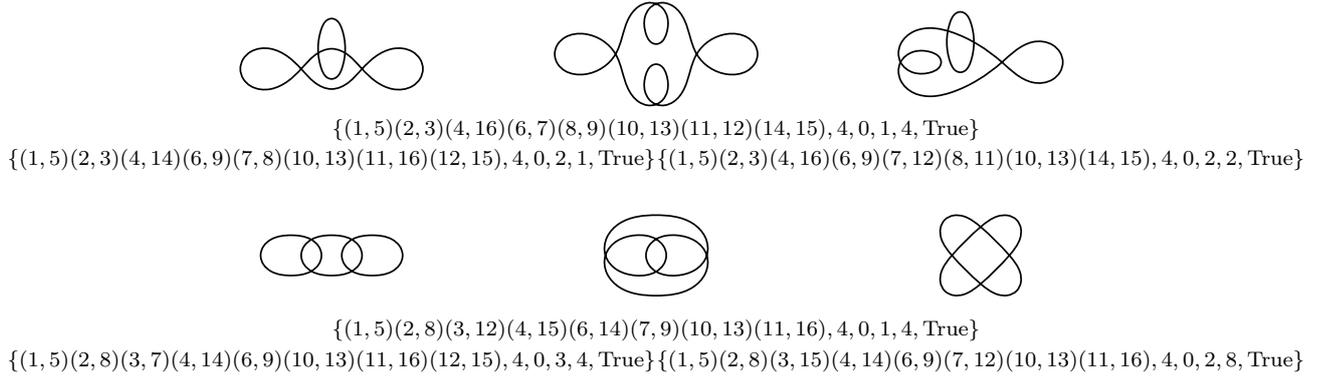
		\caption{Continuation of figure \ref{fig:example41}.}
\end{figure}

\section{Cutting graphs}\label{ch:appendixPictures}
We continue the visualization from section \ref{ch:CountingGraphs}, where we applied the boundary creation operator to the planar vacuum ribbon graphs of zeroth and first order. Each possible index of a strand has an associated line style
\begin{alignat*}{5}
	&j\quad &\tikz\draw[n, line width=\lwLegende] (0,0)--(1,0);\,, &\qquad k\quad &\tikz\draw[k, line width=\lwLegende] (0,0)--(1,0);\,, &\qquad l\quad &\tikz\draw[l, line width=\lwLegende] (0,0)--(1,0);\,, &\qquad m\quad &\tikz\draw[m, line width=\lwLegende] (0,0)--(1,0);\,, &\qquad q\quad \tikz\draw[q, line width=\lwLegende] (0,0)--(1,0);\,.
\end{alignat*}
We omit all factors of $N$, the order $|\text{Aut}(\Gamma_0)|$ of automorphism group and summation signs associated with the indices. The second graph in order $\lambda^2$ is especially interesting, because it decomposes into graphs, which have the same weight, but are actually different.\\
\begin{tikzpicture}[scale=1.85, line width=\lwMacro, font=\normalsize, miter limit=\miterLimit, baseline]

\renewcommand{\posUp}{-0.575}
\renewcommand{\posDown}{-0.15}	
	
\renewcommand{\vHeight}{0}
	\ZeroTwoOne{(\posZero,\vHeight)}{n}{m}{k}{l}
	\draw[cut] (0.1,\vHeight) -- (0.1,\vHeight+0.3) node[above] {\RNum{1}};
	\draw[cut] (0.55,\vHeight) -- (0.55,\vHeight+0.3) node[above] {\RNum{3}};
	\draw[cut] (0.9,\vHeight) -- (0.9,\vHeight+0.3) node[above] {\RNum{2}};
	\draw[cut] (0.45,\vHeight) -- (0.45,\vHeight-0.3) node[below] {\RNum{4}};
	\node[right=\eqGraphs, scale=\mathScale] at (\posZero,\vHeight) {$ \displaystyle \sim \repNKLM \frac{1}{(E_j{+}E_k)(E_j{+}E_l)^2(E_j{+}E_m)} $};
	
\renewcommand{\vHeight}{-1.2}
	\node[red,left] at (\resZero-\numGap,\vHeight) {\RNum{1}};
	\coordinate (down) at (\resZero,\vHeight-\BraceWidth);
	\coordinate (up) at (\resZero,\vHeight+\BraceWidth);
	\draw[klammer,line width=\lwBrackets] (down) node[above right] {cut $j \to$} -- (up) node[below right] {cut $k \to$};
	\scoped[scale=\smallScale] \OneTwoThree{($(up)+(\posMiniGraph,\posUp)$)}{q}{m}{l}{n};
	\scoped[scale=\smallScale] \OneTwoThree{($(down)+(\posMiniGraph,\posDown)$)}{k}{m}{l}{q};
	\coordinate (down) at (\resZero+\braceGap,\vHeight-\BraceWidth);
	\coordinate (up) at (\resZero+\braceGap,\vHeight+\BraceWidth);
	\draw[klammer,line width=\lwBrackets] (up) -- (down);
	
\renewcommand{\vHeight}{-1.2}
	\node[red,left] at (\resOne-\numGap,\vHeight) {\RNum{2}};
	\coordinate (down) at (\resOne,\vHeight-\BraceWidth);
	\coordinate (up) at (\resOne,\vHeight+\BraceWidth);
	\draw[klammer,line width=\lwBrackets] (down) node[above right] {cut $j \to$} -- (up) node[below right] {cut $m \to$};
	\scoped[scale=\smallScale] \OneTwoThree{($(up)+(\posMiniGraph,\posUp)$)}{q}{k}{l}{n};
	\scoped[scale=\smallScale] \OneTwoThree{($(down)+(\posMiniGraph,\posDown)$)}{m}{k}{l}{q};
	\coordinate (down) at (\resOne+\braceGap,\vHeight-\BraceWidth);
	\coordinate (up) at (\resOne+\braceGap,\vHeight+\BraceWidth);
	\draw[klammer,line width=\lwBrackets] (up) -- (down);
	
\renewcommand{\vHeight}{-2.2}
	\coordinate (down) at (\resOne+\braceGap,\vHeight-\BraceWidth);
	\coordinate (up) at (\resOne+\braceGap,\vHeight+\BraceWidth);
	\node[right=\eqGraphs,scale=\mathScale] at (\posZero,\vHeight) {$ \displaystyle \sim 2 \times \repNKL \frac{1}{(E_q{+}E_j)^2} \bigg(\frac{1}{(E_j{+}E_k)^2(E_j{+}E_l)} +\frac{1}{(E_q{+}E_k)^2(E_q{+}E_l)}\bigg) $};

\renewcommand{\posUp}{-0.125}
\renewcommand{\posDown}{0.425}		
	
\renewcommand{\vHeight}{-3.2}
	\node[red,left] at (\resZero-\numGap,\vHeight) {\RNum{3}};
	\coordinate (down) at (\resZero,\vHeight-\BraceWidth);
	\coordinate (up) at (\resZero,\vHeight+\BraceWidth);
	\draw[klammer,line width=\lwBrackets] (down) node[above right] {cut $j \to$} -- (up) node[below right] {cut $l \to$};
	\scoped[scale=\smallScale,shift={($(up)+(\posMiniGraph,\posUp)$)}, yscale=-1] \OneTwoOne{(0,0)}{q}{n}{k}{m};
	\scoped[scale=\smallScale,shift={($(down)+(\posMiniGraph,\posDown)$)}, yscale=-1] \OneTwoOne{(0,0)}{l}{q}{k}{m};
	\coordinate (down) at (\resZero+\braceGap,\vHeight-\BraceWidth);
	\coordinate (up) at (\resZero+\braceGap,\vHeight+\BraceWidth);
	\draw[klammer,line width=\lwBrackets] (up) -- (down);
	
\renewcommand{\vHeight}{-3.2}
	\node[red,left] at (\resOne-\numGap,\vHeight) {\RNum{4}};
	\coordinate (down) at (\resOne,\vHeight-\BraceWidth);
	\coordinate (up) at (\resOne,\vHeight+\BraceWidth);
	\draw[klammer,line width=\lwBrackets] (down) node[above right] {cut $j \to$} -- (up) node[below right] {cut $l \to$};
	\scoped[scale=\smallScale] \OneTwoOne{($(up)+(\posMiniGraph,-\posDown)$)}{q}{n}{k}{m};
	\scoped[scale=\smallScale] \OneTwoOne{($(down)+(\posMiniGraph,-\posUp)$)}{l}{q}{k}{m};
	\coordinate (down) at (\resOne+\braceGap,\vHeight-\BraceWidth);
	\coordinate (up) at (\resOne+\braceGap,\vHeight+\BraceWidth);
	\draw[klammer,line width=\lwBrackets] (up) -- (down);
	
\renewcommand{\vHeight}{-4.2}
	\coordinate (down) at (\resOne+\braceGap,\vHeight-\BraceWidth);
	\coordinate (up) at (\resOne+\braceGap,\vHeight+\BraceWidth);
	\node[right=\eqGraphs,scale=\mathScale] at (\posZero,\vHeight) {$ \displaystyle \sim 2 \times \repNKL \frac{1}{(E_q{+}E_j)^3} \bigg(\frac{1}{(E_j{+}E_k)(E_j{+}E_l)} + \frac{1}{(E_q{+}E_k)(E_q{+}E_l)} \bigg) $};

\end{tikzpicture}\\
\begin{tikzpicture}[scale=1.85, line width=\lwMacro, font=\normalsize, miter limit=\miterLimit, baseline]

\renewcommand{\posUp}{-0.35}
\renewcommand{\posDown}{0.2}

\renewcommand{\vHeight}{3.2}
	\ZeroTwoTwo{(\posZero,\vHeight)}{n}{k}{l}{m}
	\draw[cut] (0.4,\vHeight+0.05) -- (0.4,\vHeight+0.3) node[above] {\RNum{1}};
	\draw[cut] (0.8,\vHeight) -- (0.8,\vHeight+0.3) node[above] {\RNum{3}};
	\draw[cut] (0.4,\vHeight-0.05) -- (0.4,\vHeight-0.3) node[below] {\RNum{2}};
	\draw[cut] (0.4,\vHeight) -- (0.15,\vHeight) node[left] {\RNum{4}};
	\node[right=\eqGraphs, scale=\mathScale] at (\posZero,\vHeight) {$ \displaystyle \sim \repNKLM \frac{1}{(E_j{+}E_l)(E_j{+}E_k)^2(E_k{+}E_m)} $};
	
\renewcommand{\vHeight}{2}
	\node[red,left] at (\resZero-\numGap,\vHeight) {\RNum{1}};
	\coordinate (down) at (\resZero,\vHeight-\BraceWidth);
	\coordinate (up) at (\resZero,\vHeight+\BraceWidth);
	\draw[klammer,line width=\lwBrackets] (down) node[above right] {cut $l \to$} -- (up) node[below right] {cut $j \to$};
	\scoped[scale=\smallScale] \OneTwoTwo{($(up)+(\posMiniGraph,\posUp)$)}{l}{m}{q}{k};
	\scoped[scale=\smallScale] \OneTwoTwo{($(down)+(\posMiniGraph,\posDown)$)}{q}{m}{n}{k};
	\coordinate (down) at (\resZero+\braceGap,\vHeight-\BraceWidth);
	\coordinate (up) at (\resZero+\braceGap,\vHeight+\BraceWidth);
	\draw[klammer,line width=\lwBrackets] (up) -- (down);
	
\renewcommand{\vHeight}{2}
	\node[red,left] at (\resOne-\numGap,\vHeight) {\RNum{2}};
	\coordinate (down) at (\resOne,\vHeight-\BraceWidth);
	\coordinate (up) at (\resOne,\vHeight+\BraceWidth);
	\draw[klammer,line width=\lwBrackets] (down) node[above right] {cut $l \to$} -- (up) node[below right] {cut $j \to$};
	\scoped[scale=\smallScale,shift={($(up)+(\posMiniGraph,\posUp)$)}, yscale=-1] \OneTwoTwo{(0,0)}{l}{m}{q}{k};
	\scoped[scale=\smallScale,shift={($(down)+(\posMiniGraph,\posDown)$)}, yscale=-1] \OneTwoTwo{(0,0)}{q}{m}{n}{k};
	\coordinate (down) at (\resOne+\braceGap,\vHeight-\BraceWidth);
	\coordinate (up) at (\resOne+\braceGap,\vHeight+\BraceWidth);
	\draw[klammer,line width=\lwBrackets] (up) -- (down);
	
\renewcommand{\vHeight}{1}
	\coordinate (down) at (\resOne+\braceGap,\vHeight-\BraceWidth);
	\coordinate (up) at (\resOne+\braceGap,\vHeight+\BraceWidth);
	\node[right=\eqGraphs,scale=\mathScale] at (\posZero,\vHeight) {$ \displaystyle \sim 2 \times \repNKL \frac{2}{(E_q{+}E_j)^3(E_q{+}E_k)(E_j{+}E_l)} $};
	
\renewcommand{\posUp}{-0.55}
\renewcommand{\posDown}{-0.1}	
	
\renewcommand{\vHeight}{0}
	\node[red,left] at (\resZero-\numGap,\vHeight) {\RNum{3}};
	\coordinate (down) at (\resZero,\vHeight-\BraceWidth);
	\coordinate (up) at (\resZero,\vHeight+\BraceWidth);
	\draw[klammer,line width=\lwBrackets] (down) node[above right] {cut $j \to$} -- (up) node[below right] {cut $k \to$};
	\scoped[scale=\smallScale] \OneTwoFour{($(up)+(\posMiniGraph,\posUp)$)}{q}{n}{l}{m};
	\scoped[scale=\smallScale] \OneTwoFour{($(down)+(\posMiniGraph,\posDown)$)}{k}{q}{l}{m};
	\coordinate (down) at (\resZero+\braceGap,\vHeight-\BraceWidth);
	\coordinate (up) at (\resZero+\braceGap,\vHeight+\BraceWidth);
	\draw[klammer,line width=\lwBrackets] (up) -- (down);
	
\renewcommand{\vHeight}{0}
	\node[red,left] at (\resOne-\numGap,\vHeight) {\RNum{4}};
	\coordinate (down) at (\resOne,\vHeight-\BraceWidth);
	\coordinate (up) at (\resOne,\vHeight+\BraceWidth);
	\draw[klammer,line width=\lwBrackets] (down) node[above right] {cut $l$ $\to$} -- (up) node[below right] {cut $m \to$};
	\scoped[scale=\smallScale] \OneTwoFour{($(up)+(\posMiniGraph,\posUp)$)}{q}{l}{n}{k};
	\scoped[scale=\smallScale] \OneTwoFour{($(down)+(\posMiniGraph,\posDown)$)}{m}{q}{n}{k};
	\coordinate (down) at (\resOne+\braceGap,\vHeight-\BraceWidth);
	\coordinate (up) at (\resOne+\braceGap,\vHeight+\BraceWidth);
	\draw[klammer,line width=\lwBrackets] (up) -- (down);
	
\renewcommand{\vHeight}{-1}
	\coordinate (down) at (\resOne+\braceGap,\vHeight-\BraceWidth);
	\coordinate (up) at (\resOne+\braceGap,\vHeight+\BraceWidth);
	\node[right=\eqGraphs,scale=\mathScale] at (\posZero,\vHeight) {$ \displaystyle \sim 2 \times \repNKL \frac{1}{(E_q{+}E_j)^2} \bigg( \frac{1}{(E_j{+}E_k)^2(E_k{+}E_l)} + \frac{1}{(E_q{+}E_k)^2(E_k{+}E_l)} \bigg) $};

% ------------------------------------------

\renewcommand{\posUp}{-0.3}
\renewcommand{\posDown}{0.22}	
	
\renewcommand{\vHeight}{-2.1}
	\ZeroTwoThree{(\posZero,\vHeight)}{n}{k}{l}{m}
	\draw[cut] (0.8,\vHeight+0.15) -- (0.8,\vHeight+0.4) node[above] {\RNum{1}};
	\draw[cut] (0.4,\vHeight) -- (0.4,\vHeight+0.25) node[above] {\RNum{2}};
	\draw[cut] (0.6,\vHeight) -- (0.6,\vHeight-0.25) node[below] {\RNum{3}};
	\draw[cut] (0.2,\vHeight-0.15) -- (0.2,\vHeight-0.4) node[below] {\RNum{4}};
	\node[right=\eqGraphs, scale=\mathScale] at (\posZero,\vHeight) {$ \displaystyle \sim \repNKLM \frac{1}{(E_j{+}E_l)(E_l{+}E_k)(E_k{+}E_m)(E_m{+}E_j)} $};
	
\renewcommand{\vHeight}{-3.3}
	\node[red,left] at (\resZero-\numGap,\vHeight) {\RNum{1}};
	\coordinate (down) at (\resZero,\vHeight-\BraceWidth);
	\coordinate (up) at (\resZero,\vHeight+\BraceWidth);
	\draw[klammer,line width=\lwBrackets] (down) node[above right] {cut $j \to$} -- (up) node[below right] {cut $l \to$};
	\scoped[scale=\smallScale] \OneTwoFive{($(up)+(\posMiniGraph,\posUp)$)}{n}{m}{q}{k};
	\scoped[scale=\smallScale] \OneTwoFive{($(down)+(\posMiniGraph,\posDown)$)}{q}{m}{l}{k};
	\coordinate (down) at (\resZero+\braceGap,\vHeight-\BraceWidth);
	\coordinate (up) at (\resZero+\braceGap,\vHeight+\BraceWidth);
	\draw[klammer,line width=\lwBrackets] (up) -- (down);
	
\renewcommand{\vHeight}{-3.3}
	\node[red,left] at (\resOne-\numGap,\vHeight) {\RNum{2}};
	\coordinate (down) at (\resOne,\vHeight-\BraceWidth);
	\coordinate (up) at (\resOne,\vHeight+\BraceWidth);
	\draw[klammer,line width=\lwBrackets] (down) node[above right] {cut $l \to$} -- (up) node[below right] {cut $k \to$};
	\scoped[scale=\smallScale] \OneTwoFive{($(up)+(\posMiniGraph,\posUp)$)}{l}{n}{q}{m};
	\scoped[scale=\smallScale] \OneTwoFive{($(down)+(\posMiniGraph,\posDown)$)}{q}{n}{k}{m};
	\coordinate (down) at (\resOne+\braceGap,\vHeight-\BraceWidth);
	\coordinate (up) at (\resOne+\braceGap,\vHeight+\BraceWidth);
	\draw[klammer,line width=\lwBrackets] (up) -- (down);
	
\renewcommand{\vHeight}{-4.5}
	\node[red,left] at (\resZero-\numGap,\vHeight) {\RNum{3}};
	\coordinate (down) at (\resZero,\vHeight-\BraceWidth);
	\coordinate (up) at (\resZero,\vHeight+\BraceWidth);
	\draw[klammer,line width=\lwBrackets] (down) node[above right] {cut $k \to$} -- (up) node[below right] {cut $m \to$};
	\scoped[scale=\smallScale] \OneTwoFive{($(up)+(\posMiniGraph,\posUp)$)}{k}{l}{q}{n};
	\scoped[scale=\smallScale] \OneTwoFive{($(down)+(\posMiniGraph,\posDown)$)}{q}{l}{m}{n};
	\coordinate (down) at (\resZero+\braceGap,\vHeight-\BraceWidth);
	\coordinate (up) at (\resZero+\braceGap,\vHeight+\BraceWidth);
	\draw[klammer,line width=\lwBrackets] (up) -- (down);
	
\renewcommand{\vHeight}{-4.5}
	\node[red,left] at (\resOne-\numGap,\vHeight) {\RNum{4}};
	\coordinate (down) at (\resOne,\vHeight-\BraceWidth);
	\coordinate (up) at (\resOne,\vHeight+\BraceWidth);
	\draw[klammer,line width=\lwBrackets] (down) node[above right] {cut $m \to$} -- (up) node[below right] {cut $j \to$};
	\scoped[scale=\smallScale] \OneTwoFive{($(up)+(\posMiniGraph,\posUp)$)}{m}{k}{q}{l};
	\scoped[scale=\smallScale] \OneTwoFive{($(down)+(\posMiniGraph,\posDown)$)}{q}{k}{n}{l};
	\coordinate (down) at (\resOne+\braceGap,\vHeight-\BraceWidth);
	\coordinate (up) at (\resOne+\braceGap,\vHeight+\BraceWidth);
	\draw[klammer,line width=\lwBrackets] (up) -- (down);
	
\renewcommand{\vHeight}{-5.5}
	\coordinate (down) at (\resOne+\braceGap,\vHeight-\BraceWidth);
	\coordinate (up) at (\resOne+\braceGap,\vHeight+\BraceWidth);
	\node[right=\eqGraphs,scale=\mathScale] at (\posZero,\vHeight) {$ \displaystyle \sim 8 \times \repNKL \frac{1}{(E_q{+}E_j)^2(E_q{+}E_k)(E_j{+}E_l)(E_k{+}E_l)} $};	
\end{tikzpicture}\\
The sum of all weights obtained so far is equivalent to $\Omega^{(0)}_q$ up to second order. To obtain the graphs that contribute to $\Omega^{(0)}_{q,r}$, we apply $\hat{T}_r$ with $r\neq q$ to the graphs above. Instead of considering every ribbon individually, we will group the cuts of ribbons with the same indices together. Every cut with number \RNum{1} will produce disconnected graphs. \\
\begin{tikzpicture}[scale=1.85, line width=\lwMacro, font=\normalsize, miter limit=\miterLimit]
	
\renewcommand{\vHeight}{0}
	\OneZero{(\posZero,\vHeight)}{q}{n}
	\node[right=\eqGraphs, scale=\mathScale] at (\posZero,\vHeight) {$ \displaystyle \sim \repN \frac{1}{E_q{+}E_j} $};
	\draw[cut] (\posZero+0.5,\vHeight-0.15) -- (\posZero+0.5,\vHeight+0.15) node[above] {\RNum{1}};
	
\renewcommand{\vHeight}{0}
	\node[red,left] at (\resOne-\numGap,\vHeight) {\RNum{1}};
	\coordinate (down) at (\resOne,\vHeight-\BraceSizeOne);
	\coordinate (up) at (\resOne,\vHeight+\BraceSizeOne);
	\draw[klammer,line width=\lwBrackets] (down) -- (up) node[pos=0.5, right] (middleCut) {cut $j \to$};
	\node[right=\eqArrow] at (middleCut) {$ \displaystyle  \frac{1}{(E_q{+}E_r)^2} $};
\end{tikzpicture}\\
\begin{tikzpicture}[scale=1.85, line width=\lwMacro, font=\normalsize, miter limit=\miterLimit]
\renewcommand{\vHeight}{1.9}
	\OneOne{(\posZero,\vHeight)}{q}{n}{k}
	\OneOne{(\posOne,\vHeight)}{n}{q}{k}
	\node[scale=\timesScale] at ($(\posZero+0.5,\vHeight)!0.5!(\posOne+0.5,\vHeight)$) {+};
	\node[right=\eqGraphs, scale=\mathScale] at (\posOne,\vHeight) {$ \displaystyle \sim \repNK \frac{1}{(E_q+E_j)^2} \bigg(\frac{1}{E_j{+}E_k} + \frac{1}{E_q{+}E_k} \bigg) $};
	\draw[cut] (\posZero+0.1,\vHeight-0.15) -- (\posZero+0.1,\vHeight+0.15) node[above] {\RNum{1}};
	\draw[cut] (\posZero+0.9,\vHeight-0.15) -- (\posZero+0.9,\vHeight+0.15) node[above] {\RNum{1}};
	\draw[cut] (\posOne+0.1,\vHeight-0.15) -- (\posOne+0.1,\vHeight+0.15) node[above] {\RNum{1}};
	\draw[cut] (\posOne+0.9,\vHeight-0.15) -- (\posOne+0.9,\vHeight+0.15) node[above] {\RNum{1}};
	
	\draw[cut] (\posZero+0.5,\vHeight+0.4) -- (\posZero+0.8,\vHeight+0.6) node[right] {\RNum{2}};
	\draw[cut] (\posOne+0.5,\vHeight+0.4) -- (\posOne+0.8,\vHeight+0.6) node[right] {\RNum{2}};
	
\renewcommand{\vHeight}{1}
	\node[red,left] at (\resZero-\numGap,\vHeight) {\RNum{1}};
	\coordinate (down) at (\resZero,\vHeight-\BraceSizeOne);
	\coordinate (up) at (\resZero,\vHeight+\BraceSizeOne);
	\draw[klammer,line width=\lwBrackets] (down) -- (up) node[pos=0.5, right] (middleCut) {cut $j \to$};
	\node[right=\eqArrow] at (middleCut) {$ \displaystyle  \repN \frac{2}{(E_q{+}E_r)^3(E_r{+}E_j)} + \frac{2}{(E_q{+}E_r)^3(E_q{+}E_j)} $};	
	
\renewcommand{\vHeight}{0}
	\node[red,left] at (\resZero-\numGap,\vHeight) {\RNum{2}};
	\coordinate (down) at (\resZero,\vHeight-\BraceSizeTwo);
	\coordinate (up) at (\resZero,\vHeight+\BraceSizeTwo);
	\draw[klammer,line width=\lwBrackets] (down) -- (up) node[pos=\CutPos, right] (downCut) {cut $k \to$} node[pos=1-\CutPos,right] (upCut) {cut $j \to$};
	\node[right=\eqArrow] at (upCut) {$ \displaystyle  \repN \frac{1}{(E_q{+}E_r)^2(E_r{+}E_j)^2}$};
	\node[right=\eqArrow] at (downCut) {$ \displaystyle  \repN \frac{1}{(E_q{+}E_j)^2(E_r{+}E_j)^2} + \frac{1}{(E_q{+}E_j)^2(E_q{+}E_r)^2} $};
	
% ------------------------------------------

\renewcommand{\vHeight}{-1.8}
	\OneTwoThree{(\posZero,\vHeight)}{q}{l}{k}{n}
	\OneTwoThree{(\posOne,\vHeight)}{n}{l}{k}{q}
	\node[scale=\timesScale] at ($(\posZero+0.5,\vHeight)!0.5!(\posOne+0.5,\vHeight)$) {+};
	\node[right=\eqGraphs, scale=\mathScale] at (\posOne,\vHeight) {$ \displaystyle \sim \repNKL \frac{1}{(E_q{+}E_j)^2} \bigg(\frac{1}{(E_j{+}E_k)^2(E_j{+}E_l)} +\frac{1}{(E_q{+}E_k)^2(E_q{+}E_l)}\bigg) $};
	\draw[cut] (\posZero+0.2,\vHeight-0.15) node[below] {\RNum{1}} -- (\posZero+0.2,\vHeight+0.15);
	\draw[cut] (\posZero+0.8,\vHeight-0.15) node[below] {\RNum{1}} -- (\posZero+0.8,\vHeight+0.15);
	\draw[cut] (\posOne+0.2,\vHeight-0.15) node[below] {\RNum{1}} -- (\posOne+0.2,\vHeight+0.15);
	\draw[cut] (\posOne+0.8,\vHeight-0.15) node[below] {\RNum{1}} -- (\posOne+0.8,\vHeight+0.15);
	
	\draw[cut] (\posZero+0.5,\vHeight+0.325) -- (\posZero+0.2,\vHeight+0.325) node[left] {\RNum{2}};
	\draw[cut] (\posZero+0.5,\vHeight+0.225) -- (\posZero+0.8,\vHeight+0.225) node[right] {\RNum{2}};
	\draw[cut] (\posOne+0.5,\vHeight+0.325) -- (\posOne+0.2,\vHeight+0.325) node[left] {\RNum{2}};
	\draw[cut] (\posOne+0.5,\vHeight+0.225) -- (\posOne+0.8,\vHeight+0.225) node[right] {\RNum{2}};

	\draw[cut] (\posZero+0.5,\vHeight+0.65) -- (\posZero+0.8,\vHeight+0.65) node[right] {\RNum{3}};
	\draw[cut] (\posOne+0.5,\vHeight+0.65) -- (\posOne+0.8,\vHeight+0.65) node[right] {\RNum{3}};
	
\renewcommand{\vHeight}{-2.75}
	\node[red,left] at (\resZero-\numGap,\vHeight) {\RNum{1}};
	\coordinate (down) at (\resZero,\vHeight-\BraceSizeOne);
	\coordinate (up) at (\resZero,\vHeight+\BraceSizeOne);
	\draw[klammer,line width=\lwBrackets] (down) -- (up) node[pos=0.5, right] (middleCut) {cut $j \to$};
	\node[right=\eqArrow] at (middleCut) {$ \displaystyle  \repNK \frac{2}{(E_q{+}E_r)^3(E_r{+}E_j)^2(E_r{+}E_k)} + \frac{2}{(E_q{+}E_r)^3(E_q{+}E_j)^2(E_q{+}E_k)} $};
	
\renewcommand{\vHeight}{-3.7}
	\node[red,left] at (\resZero-\numGap,\vHeight) {\RNum{2}};
	\coordinate (down) at (\resZero,\vHeight-\BraceSizeTwo);
	\coordinate (up) at (\resZero,\vHeight+\BraceSizeTwo);
	\draw[klammer,line width=\lwBrackets] (down) -- (up) node[pos=\CutPos, right] (downCut) {cut $k \to$} node[pos=1-\CutPos,right] (upCut) {cut $j \to$};
	\node[right=\eqArrow] at (upCut) {$ \displaystyle  \repNK \frac{2}{(E_q{+}E_r)^2(E_r{+}E_j)^3(E_r{+}E_k)}$};
	\node[right=\eqArrow] at (downCut) {$ \displaystyle  \repNK \frac{2}{(E_q{+}E_j)^2(E_r{+}E_j)^3(E_j{+}E_k)} + \frac{2}{(E_q{+}E_j)^2(E_q{+}E_r)^3(E_q{+}E_k)} $};
	
\renewcommand{\vHeight}{-4.9}
	\node[red,left] at (\resZero-\numGap,\vHeight) {\RNum{3}};
	\coordinate (down) at (\resZero,\vHeight-\BraceSizeTwo);
	\coordinate (up) at (\resZero,\vHeight+\BraceSizeTwo);
	\draw[klammer,line width=\lwBrackets] (down) -- (up) node[pos=\CutPos, right] (downCut) {cut $l \to$} node[pos=1-\CutPos,right] (upCut) {cut $j \to$};
	\node[right=\eqArrow] at (upCut) {$ \displaystyle  \repNK \frac{1}{(E_q{+}E_r)^2(E_r{+}E_j)^2(E_r{+}E_k)^2}$};
	\node[right=\eqArrow] at (downCut) {$ \displaystyle  \repNK \frac{1}{(E_q{+}E_j)^2(E_j{+}E_k)^2(E_j{+}E_r)^2} + \frac{1}{(E_q{+}E_j)^2(E_q{+}E_k)^2(E_q{+}E_r)^2} $};	
	
% ------------------------------------------
	
\renewcommand{\vHeight}{-6.7}
	\OneTwoOne{(\posZero,\vHeight)}{q}{n}{k}{l}
	\OneTwoOne{(\posOne,\vHeight)}{n}{q}{k}{l}
	\node[scale=\timesScale] at ($(\posZero+0.5,\vHeight)!0.5!(\posOne+0.5,\vHeight)$) {+};
	\node[right=\eqGraphs, scale=\mathScale] at (\posOne,\vHeight) {$ \displaystyle \sim \repNKL \frac{1}{(E_q{+}E_j)^3} \bigg(\frac{1}{(E_j{+}E_k)(E_j{+}E_l)} + \frac{1}{(E_q{+}E_k)(E_q{+}E_l)} \bigg) $};
	\draw[cut] (\posZero+0.05,\vHeight-0.15) node[below] {\RNum{1}} -- (\posZero+0.05,\vHeight+0.15);
	\draw[cut] (\posZero+0.95,\vHeight-0.15) node[below] {\RNum{1}} -- (\posZero+0.95,\vHeight+0.15);
	\draw[cut] (\posOne+0.05,\vHeight-0.15) node[below] {\RNum{1}} -- (\posOne+0.05,\vHeight+0.15);
	\draw[cut] (\posOne+0.95,\vHeight-0.15) node[below] {\RNum{1}} -- (\posOne+0.95,\vHeight+0.15);
	
	\draw[cut] (\posZero+0.275,\vHeight+0.25) -- (\posZero+0.275,\vHeight+0.5) node[above] {\RNum{2}};
	\draw[cut] (\posZero+0.725,\vHeight+0.25) -- (\posZero+0.725,\vHeight+0.5) node[above] {\RNum{2}};
	\draw[cut] (\posOne+0.275,\vHeight+0.25) -- (\posOne+0.275,\vHeight+0.5) node[above] {\RNum{2}};
	\draw[cut] (\posOne+0.725,\vHeight+0.25) -- (\posOne+0.725,\vHeight+0.5) node[above] {\RNum{2}};
	
\renewcommand{\vHeight}{-7.6}
	\node[red,left] at (\resZero-\numGap,\vHeight) {\RNum{1}};
	\coordinate (down) at (\resZero,\vHeight-\BraceSizeOne);
	\coordinate (up) at (\resZero,\vHeight+\BraceSizeOne);
	\draw[klammer,line width=\lwBrackets] (down) -- (up) node[pos=0.5, right] (middleCut) {cut $j \to$};
	\node[right=\eqArrow] at (middleCut) {$ \displaystyle  \repNK \frac{3}{(E_q{+}E_r)^4(E_r{+}E_j)(E_r{+}E_k)} + \frac{3}{(E_q{+}E_r)^4(E_q{+}E_j)(E_q{+}E_k)} $};
	
\renewcommand{\vHeight}{-8.85}
	\node[red,left] at (\resZero-\numGap,\vHeight) {\RNum{2}};
	\coordinate (down) at (\resZero,\vHeight-\BraceSizeThree);
	\coordinate (up) at (\resZero,\vHeight+\BraceSizeThree);
	\draw[klammer,line width=\lwBrackets] (down) -- (up) node[pos=\CutPosBig, right] (downCut) {cut $l \to$} node[pos=0.5, right] (middleCut) {cut $k \to$} node[pos=1-\CutPosBig,right] (upCut) {cut $j \to$};
	\node[right=\eqArrow] at (upCut) {$ \displaystyle  \repNK 2 \times \frac{1}{(E_q{+}E_r)^3(E_r{+}E_j)^2(E_r{+}E_k)}$};
	\node[right=\eqArrow] at (middleCut) {$ \displaystyle  \repNK \frac{1}{(E_q{+}E_j)^3(E_r{+}E_j)^2(E_j{+}E_k)} + \frac{1}{(E_q{+}E_j)^3(E_q{+}E_r)^2(E_q{+}E_k)}$};
	\node[right=\eqArrow] at (downCut) {$ \displaystyle  \repNK \frac{1}{(E_q{+}E_j)^3(E_j{+}E_k)(E_r{+}E_j)^2} + \frac{1}{(E_q{+}E_j)^3(E_q{+}E_k)(E_q{+}E_r)^2} $};
\end{tikzpicture}

\begin{tikzpicture}[scale=1.85, line width=\lwMacro, font=\normalsize, miter limit=\miterLimit]

% ------------------------------------------
	
\renewcommand{\vHeight}{0}
	
	\OneTwoTwo{(\posZero,\vHeight)}{q}{k}{n}{l}
	\scoped[shift={((\posOne,\vHeight)}, yscale=-1] \OneTwoTwo{(0,0)}{n}{l}{q}{k};
	\node[scale=\timesScale] at ($(\posZero+0.5,\vHeight)!0.5!(\posOne+0.5,\vHeight)$) {+};
	\node[right=\eqGraphs, scale=\mathScale] at (\posOne,\vHeight) {$ \displaystyle \sim \repNKL \frac{2}{(E_q{+}E_j)^3(E_q{+}E_k)(E_j{+}E_l)} $};
	\draw[cut] (\posZero+0.05,\vHeight-0.15) -- (\posZero+0.05,\vHeight+0.15) node[above] {\RNum{1}};
	\draw[cut] (\posZero+0.95,\vHeight-0.15) -- (\posZero+0.95,\vHeight+0.15) node[above] {\RNum{1}};
	\draw[cut] (\posZero+0.5,\vHeight-0.15) -- (\posZero+0.5,\vHeight+0.15) node[above] {\RNum{1}};
	
	\draw[cut] (\posZero+0.275,\vHeight+0.25) -- (\posZero+0.275,\vHeight+0.5) node[above] {\RNum{2}};
	\draw[cut] (\posZero+0.725,\vHeight-0.25) -- (\posZero+0.725,\vHeight-0.5) node[below] {\RNum{2}};
	
	\draw[cut] (\posOne+0.05,\vHeight-0.15) -- (\posOne+0.05,\vHeight+0.15) node[above] {\RNum{1}};
	\draw[cut] (\posOne+0.95,\vHeight-0.15) -- (\posOne+0.95,\vHeight+0.15) node[above] {\RNum{1}};
	\draw[cut] (\posOne+0.5,\vHeight-0.15) -- (\posOne+0.5,\vHeight+0.15) node[above] {\RNum{1}};
	
	\draw[cut] (\posOne+0.725,\vHeight+0.25) -- (\posOne+0.725,\vHeight+0.5) node[above] {\RNum{2}};
	\draw[cut] (\posOne+0.275,\vHeight-0.25) -- (\posOne+0.275,\vHeight-0.5) node[below] {\RNum{2}};
	
\renewcommand{\vHeight}{-1.2}
	\node[red,left] at (\resZero-\numGap,\vHeight) {\RNum{1}};
	\coordinate (down) at (\resZero,\vHeight-\BraceSizeOne);
	\coordinate (up) at (\resZero,\vHeight+\BraceSizeOne);
	\draw[klammer,line width=\lwBrackets] (down) -- (up) node[pos=0.5, right] (middleCut) {cut $j \to$};
	\node[right=\eqArrow] at (middleCut) {$ \displaystyle  \repNK 2 \times \frac{3}{(E_q{+}E_r)^4(E_q{+}E_j)(E_r{+}E_k)} $};
	
\renewcommand{\vHeight}{-2.45}
	\node[red,left] at (\resZero-\numGap,\vHeight) {\RNum{2}};
	\coordinate (down) at (\resZero,\vHeight-\BraceSizeThree);
	\coordinate (up) at (\resZero,\vHeight+\BraceSizeThree);
	\draw[klammer,line width=\lwBrackets] (down) -- (up) node[pos=\CutPosBig, right] (downCut) {cut $l \to$} node[pos=0.5, right] (middleCut) {cut $k \to$} node[pos=1-\CutPosBig,right] (upCut) {cut $j \to$};
	\node[right=\eqArrow] at (upCut) {$ \displaystyle  \repNK 2 \times \frac{1}{(E_q{+}E_r)^3(E_q{+}E_k)(E_r{+}E_j)^2}$};
	\node[right=\eqArrow] at (middleCut) {$ \displaystyle  \repNK 2 \times \frac{1}{(E_q{+}E_j)^3(E_q{+}E_r)^2(E_j{+}E_k)} $};
	\node[right=\eqArrow] at (downCut) {$ \displaystyle  \repNK 2 \times \frac{1}{(E_q{+}E_j)^3(E_q{+}E_k)(E_r{+}E_j)^2} $};
	
% ------------------------------------------

\renewcommand{\vHeight}{-4.3}
	\OneTwoFour{(\posZero,\vHeight)}{q}{n}{k}{l}
	\OneTwoFour{(\posOne,\vHeight)}{n}{q}{k}{l}
	\node[scale=\timesScale] at ($(\posZero+0.5,\vHeight)!0.5!(\posOne+0.5,\vHeight)$) {+};
	\node[right=\eqGraphs, scale=\mathScale] at (\posOne,\vHeight) {$ \displaystyle \sim \repNKL \frac{1}{(E_q{+}E_j)^2} \bigg( \frac{1}{(E_j{+}E_k)^2(E_k{+}E_l)} + \frac{1}{(E_q{+}E_k)^2(E_k{+}E_l)} \bigg) $};
	\draw[cut] (\posZero+0.2,\vHeight-0.15) node[below] {\RNum{1}} -- (\posZero+0.2,\vHeight+0.15);
	\draw[cut] (\posZero+0.8,\vHeight-0.15) node[below] {\RNum{1}} -- (\posZero+0.8,\vHeight+0.15);
	\draw[cut] (\posOne+0.2,\vHeight-0.15) node[below] {\RNum{1}} -- (\posOne+0.2,\vHeight+0.15);
	\draw[cut] (\posOne+0.8,\vHeight-0.15) node[below] {\RNum{1}} -- (\posOne+0.8,\vHeight+0.15);
	
	\draw[cut] (\posZero+0.4,\vHeight+0.25) -- (\posZero+0.1,\vHeight+0.25) node[left] {\RNum{2}};
	\draw[cut] (\posZero+0.6,\vHeight+0.25) -- (\posZero+0.9,\vHeight+0.25) node[right] {\RNum{2}};
	\draw[cut] (\posOne+0.4,\vHeight+0.25) -- (\posOne+0.1,\vHeight+0.25) node[left] {\RNum{2}};
	\draw[cut] (\posOne+0.6,\vHeight+0.25) -- (\posOne+0.9,\vHeight+0.25) node[right] {\RNum{2}};
	
	\draw[cut] (\posZero+0.5,\vHeight+0.2) node[below] {\RNum{3}} -- (\posZero+0.5,\vHeight+0.4);
	\draw[cut] (\posOne+0.5,\vHeight+0.2) node[below] {\RNum{3}} -- (\posOne+0.5,\vHeight+0.4);
	
\renewcommand{\vHeight}{-5.2}
	\node[red,left] at (\resZero-\numGap,\vHeight) {\RNum{1}};
	\coordinate (down) at (\resZero,\vHeight-\BraceSizeOne);
	\coordinate (up) at (\resZero,\vHeight+\BraceSizeOne);
	\draw[klammer,line width=\lwBrackets] (down) -- (up) node[pos=0.5, right] (middleCut) {cut $j \to$};
	\node[right=\eqArrow] at (middleCut) {$ \displaystyle  \repNK \frac{2}{(E_q{+}E_r)^3(E_r{+}E_j)^2(E_j{+}E_k)} + \frac{2}{(E_q{+}E_r)^3(E_q{+}E_j)^2(E_j{+}E_k)} $};
	
\renewcommand{\vHeight}{-6.2}
	\node[red,left] at (\resZero-\numGap,\vHeight) {\RNum{2}};
	\coordinate (down) at (\resZero,\vHeight-\BraceSizeTwo);
	\coordinate (up) at (\resZero,\vHeight+\BraceSizeTwo);
	\draw[klammer,line width=\lwBrackets] (down) -- (up) node[pos=\CutPos, right] (downCut) {cut $k \to$} node[pos=1-\CutPos,right] (upCut) {cut $j \to$};
	\node[right=\eqArrow] at (upCut) {$ \displaystyle  \repNK \frac{2}{(E_q{+}E_r)^2(E_r{+}E_j)^3(E_j{+}E_k)}$};
	\node[right=\eqArrow] at (downCut) {$ \displaystyle  \repNK \frac{2}{(E_q{+}E_j)^2(E_r{+}E_j)^3(E_r{+}E_k)} + \frac{2}{(E_q{+}E_j)^2(E_q{+}E_r)^3(E_r{+}E_k)} $};
	
\renewcommand{\vHeight}{-7.45}
	\node[red,left] at (\resZero-\numGap,\vHeight) {\RNum{3}};
	\coordinate (down) at (\resZero,\vHeight-\BraceSizeTwo);
	\coordinate (up) at (\resZero,\vHeight+\BraceSizeTwo);
	\draw[klammer,line width=\lwBrackets] (down) -- (up) node[pos=\CutPos, right] (downCut) {cut $l \to$} node[pos=1-\CutPos,right] (upCut) {cut $k \to$};
	\node[right=\eqArrow] at (upCut) {$ \displaystyle  \repNK \frac{1}{(E_q{+}E_j)^2(E_r{+}E_j)^2(E_r{+}E_k)^2} + \frac{1}{(E_q{+}E_j)^2(E_q{+}E_r)^2(E_r{+}E_k)^2}$};
	\node[right=\eqArrow] at (downCut) {$ \displaystyle  \repNK \frac{1}{(E_q{+}E_j)^2(E_j{+}E_k)^2(E_r{+}E_k)^2} + \frac{1}{(E_q{+}E_j)^2(E_q{+}E_k)^2(E_r{+}E_k)^2} $};
% ------------------------------------------
	
\renewcommand{\vHeight}{-8.8}
	\OneTwoFive{(\posZero,\vHeight)}{n}{l}{q}{k}
	\node[right=\eqGraphs, scale=\mathScale] at (\posZero,\vHeight) {$ \displaystyle \sim \repNKL \frac{1}{(E_q{+}E_j)^2(E_q{+}E_k)(E_j{+}E_l)(E_k{+}E_l)} $};
	\draw[cut] (\posZero+0.1,\vHeight-0.15) -- (\posZero+0.1,\vHeight+0.15) node[above] {\RNum{1}};
	\draw[cut] (\posZero+0.9,\vHeight-0.15) -- (\posZero+0.9,\vHeight+0.15) node[above] {\RNum{1}};
	
	\draw[cut] (\posZero+0.5,\vHeight+0.15) -- (\posZero+0.5,\vHeight+0.35) node[above] {\RNum{2}};
	
	\draw[cut] (\posZero+0.5,\vHeight-0.1) -- (\posZero+0.5,\vHeight+0.1) node[midway,right] {\RNum{3}};
	
	\draw[cut] (\posZero+0.5,\vHeight-0.15) -- (\posZero+0.5,\vHeight-0.35) node[below] {\RNum{4}};
	
\renewcommand{\vHeight}{-9.8}
	\node[red,left] at (\resZero-\numGap,\vHeight) {\RNum{1}};
	\coordinate (down) at (\resZero,\vHeight-\BraceSizeOne);
	\coordinate (up) at (\resZero,\vHeight+\BraceSizeOne);
	\draw[klammer,line width=\lwBrackets] (down) -- (up) node[pos=0.5, right] (middleCut) {cut $j \to$};
	\node[right=\eqArrow] at (middleCut) {$ \displaystyle  \repNK \frac{2}{(E_q{+}E_r)^3(E_q{+}E_j)(E_r{+}E_k)(E_j{+}E_k)} $};
	
\renewcommand{\vHeight}{-10.55}
	\node[red,left] at (\resZero-\numGap,\vHeight) {\RNum{2}};
	\coordinate (down) at (\resZero,\vHeight-\BraceSizeOne);
	\coordinate (up) at (\resZero,\vHeight+\BraceSizeOne);
	\draw[klammer,line width=\lwBrackets] (down) -- (up) node[pos=0.5, right] (middleCut) {cut $k \to$};
	\node[right=\eqArrow] at (middleCut) {$ \displaystyle  \repNK \frac{1}{(E_q{+}E_j)^2(E_q{+}E_r^2)(E_j{+}E_k)(E_r{+}E_k)} $};

\renewcommand{\vHeight}{-11.55}
	\node[red,left] at (\resZero-\numGap,\vHeight) {\RNum{3}};
	\coordinate (down) at (\resZero,\vHeight-\BraceSizeTwo);
	\coordinate (up) at (\resZero,\vHeight+\BraceSizeTwo);
	\draw[klammer,line width=\lwBrackets] (down) -- (up) node[pos=\CutPos, right] (downCut) {cut $l \to$} node[pos=1-\CutPos,right] (upCut) {cut $j \to$};
	\node[right=\eqArrow] at (upCut) {$ \displaystyle  \repNK \frac{1}{(E_q{+}E_r)^2(E_q{+}E_j)(E_r{+}E_k)^2(E_j{+}E_k)} $};
	\node[right=\eqArrow] at (downCut) {$ \displaystyle  \repNK \frac{1}{(E_q{+}E_j)^2(E_q{+}E_k)(E_r{+}E_j)^2(E_r{+}E_k)} $};
	
\end{tikzpicture}\newpage
\begin{tikzpicture}[scale=2, line width=\lwMacro, font=\normalsize, miter limit=\miterLimit]
	
\renewcommand{\vHeight}{-12.5}
	\node[red,left] at (\resZero-\numGap,\vHeight) {\RNum{4}};
	\coordinate (down) at (\resZero,\vHeight-\BraceSizeTwo);
	\coordinate (up) at (\resZero,\vHeight+\BraceSizeTwo);
	\draw[klammer,line width=\lwBrackets] (down) -- (up) node[pos=\CutPos, right] (downCut) {cut $l \to$} node[pos=1-\CutPos,right] (upCut) {cut $k \to$};
	\node[right=\eqArrow] at (upCut) {$ \displaystyle  \repNK \frac{1}{(E_q{+}E_j)^2(E_q{+}E_r)(E_j{+}E_k)(E_r{+}E_k)^2} $};
	\node[right=\eqArrow] at (downCut) {$ \displaystyle  \repNK \frac{1}{(E_q{+}E_j)^2(E_q{+}E_k)(E_r{+}E_j)(E_r{+}E_k)^2} $};
\end{tikzpicture}

\bibliographystyle{halpha-abbrv}
\bibliography{references}

\end{document}